\documentclass[reprint,aps,prc,amsmath,amssymb,nofootinbib,superscriptaddress]{revtex4-1}
  
\usepackage[utf8]{inputenc}
\usepackage[T1]{fontenc}
\usepackage[normalem]{ulem}
\usepackage{tabularx}
\usepackage{balance}
\usepackage{amsmath}
\usepackage{mathtools}
\usepackage{braket}
\usepackage{tensor}
\usepackage{xcolor}
\usepackage[abbreviations]{siunitx}
\usepackage{booktabs}
\usepackage{float}
\usepackage{placeins}
\usepackage{graphicx,color,rotating,pifont}
\usepackage{siunitx}
\usepackage{bm}
\usepackage{ae}
\usepackage{dcolumn}
\usepackage{txfonts}
\usepackage{tensor}
\usepackage{booktabs}
\usepackage{xspace}
\usepackage{mathrsfs}
\usepackage{amssymb} 
\usepackage{amsmath}
\usepackage{esvect}
\usepackage{diagbox}
\usepackage{soul}
\setstcolor{red}
\setul{}{.2ex}

\usepackage{tikz}
\tikzset{asg/.cd,
  omega-vertex/.style={circle,solid,draw=black,fill=white,minimum size=5pt, inner sep=0pt},
  dbd-vertex/.style={coordinate},
  pline/.style={thick, postaction={decorate}, decoration={markings, mark=at position .5 with {\arrow[xshift=2pt]{stealth}}}},
  hline/.style={thick, postaction={decorate}, decoration={markings, mark=at position .5 with {\arrowreversed[xshift=-2pt]{stealth}}}},
  shift arrow/.style={/pgf/decoration/transform={xshift=#1}},
  shift arrow/.default=-2pt,
  dbd-2b/.style={decorate, decoration=snake},
  omega-2b/.style={densely dashed},
  neutron/.style={draw=blue},
  proton/.style={draw=red},
}

\usepackage[colorlinks,linkcolor=blue,anchorcolor=blue,citecolor=blue]{hyperref} 
\hypersetup{
   colorlinks=true, 
   linkcolor=blue,  
   citecolor=blue,  
   filecolor=blue,  
   urlcolor=blue    
}

\DeclareMathSymbol{\NS}{\mathord}{AMSb}{"4E}

\DeclareSIUnit{\fm}{\femto\meter}

\newcommand{\asurf}{a_{\mathrm{surf}}}

\newcommand{\assym}{a_{\mathrm{ssym}}}
\newcommand{\bt}{\beta_2}

\newcommand{\nuc}[2]{\ensuremath{{}^{#2}\mathrm{#1}}}

\newcommand{\beq}{\begin{equation}}
\newcommand{\eeq}{\end{equation}}
\newcommand{\beqn}{\begin{eqnarray}}
\newcommand{\eeqn}{\end{eqnarray}}
\newcommand{\bsub}{\begin{subequations}}
\newcommand{\esub}{\end{subequations}}
\newcommand{\bpm}{\begin{pmatrix}}
\newcommand{\epm}{\end{pmatrix}}

\newcommand\identity{1\kern-0.25em\text{l}}

\renewcommand{\vec}[1]{\mbox{\boldmath $#1$}}

\begin{document} 
\title{Sensitivity of fission barrier heights to nuclear matter properties}
 
 \author{X. Zhang}  
 \email{zhang.xin.38d@st.kyoto-u.ac.jp}
    \affiliation{Department of Physics, Kyoto University,
Kyoto 606-8502, Japan}    
  
   \author{K. Hagino} 
   \email{hagino.kouichi.5m@kyoto-u.ac.jp}
    \affiliation{Department of Physics, Kyoto University,
Kyoto 606-8502, Japan}  
  \affiliation{Institute for Liberal Arts and Sciences, Kyoto University, Kyoto 606-8501, Japan}
\affiliation{ 
RIKEN Nishina Center for Accelerator-based Science, RIKEN, Wako 351-0198, Japan
}
\author{J. M. Yao}
\email{yaojm8@sysu.edu.cn}
\affiliation{School of Physics and Astronomy, Sun Yat-sen University, Zhuhai 519082, P.R. China}   
\affiliation{Yukawa Institute for Theoretical Physics, Kyoto University, Kyoto, 606-8502, Japan}   

\begin{abstract} 
We systematically investigate the correlations between nuclear fission properties and the surface energy coefficient $a_{\rm surf}$ extracted from semi-infinite nuclear matter. To this end,  we employ the covariant density functional theory based on a point-coupling energy density functional (EDF). Taking \nuc{Pu}{240} and \nuc{U}{236} as representative examples, we employ Gaussian-process emulators to construct the joint posterior distributions of $a_{\rm surf}$ and selected fission observables, constrained by the ground state properties of the two nuclei. The results reveal a strong correlation between $a_{\rm surf}$ and the fission barrier heights. A sensitivity analysis further shows that the height of the outer fission barrier is governed predominantly by the isoscalar-vector derivative coupling constant $\delta_V$ in the point-coupling EDF, whereas that of the inner barrier is by the combined effects of $\delta_V$ and several mutually correlated coupling constants. 

\end{abstract}
  
\maketitle

\section{Introduction}
Nuclear fission plays a central role in a broad range of phenomena in nuclear physics~\cite{Krappe2012,Schunck2016,Andreyev2018}, such as nuclear energy production~\cite{Schmidt2018}, the laboratory synthesis of superheavy elements~\cite{Oganessian2015,Hofmann2015}, and the $r$-process nucleosynthesis of heavy elements in astrophysical environments~\cite{MUMPOWER2016,Giuliani2018,cowan2021heaviest}. A key quantity characterizing the fission process is the fission barrier height, defined as the energy difference between the nuclear ground state and the relevant saddle point along the energetically favored fission path on the multidimensional, shape-constrained potential-energy surface (PES)~\cite{Bertsch2015}. An accurate microscopic description of fission barriers is therefore essential for reliable predictions of fission observables. At the same time, fission barriers provide one of the most stringent tests of nuclear structure theory, as their description requires a consistent treatment of large-amplitude collective deformation together with the evolution of the underlying single-particle shell structure over a wide range of nuclear shapes~\cite{Bender2003,Nakatsukasa2016,SCHUNCK2022}.

Nuclear density functional theory (DFT) \cite{Bender2003,Vretenar:2005PR} has emerged as one of the most successful microscopic frameworks for describing nuclear fission across the nuclear chart~\cite{Schmidt2018,Bender:2020oze}. A broad range of energy density functionals (EDFs), including nonrelativistic Skyrme and Gogny functionals, as well as covariant formulation, have been employed to investigate fission barrier heights, fission isomers, and spontaneous-fission half-lives~\cite{Guzman2014,taninah2020covariant,Washiyama2021,Washiyama2024}. Moreover, DFT provides a unified description of finite nuclei and infinite nuclear matter~\cite{Dutra2014,Yuksel2019,Yuksel2021}, thereby offering a natural framework for systematically exploring the connections between bulk nuclear matter properties and observables of finite nuclei~\cite{Mondal2015,Agrawal2021,An2023MirrorRadius}. Prominent examples of such connections include the well-established correlation between neutron-skin thickness and the density dependence of the symmetry energy~\cite{Brown2000,Centelles2009}, as well as that between the giant monopole resonance energy and nuclear incompressibility~\cite{BLAIZOT1980,GARG201855}. By comparison, considerably less attention has been devoted to understanding how nuclear matter properties influence fission observables, particularly fission barrier heights~\cite{SEIF201877,Ryssens2019}.

It is well established that nuclear fission is governed to a large extent by the competition between the surface and the Coulomb energies, with microscopic shell effects providing an additional contribution that is crucial for determining the detailed structure and height of the fission barrier. This suggests that fission properties may exhibit a pronounced correlation with the surface energy. In the liquid drop model~\cite{MYERS1980267}, the surface energy is characterized by the coefficient $a_{\rm surf}$, which controls the energetic cost associated with the increase of nuclear surface area during deformation. Unlike bulk nuclear matter properties, which are defined in infinite nuclear matter (INM), the surface energy coefficient $a_{\rm surf}$ characterizes the surface properties of semi-infinite nuclear matter (SINM)~\cite{CENTELLES1998193,Patra2002,Reinhard2006,Jodon2016}. An important investigation of the connection between $a_{\rm surf}$ and fission properties has been reported in Ref.~\cite{Ryssens2019}. In that work, the SLy5sX family of Skyrme parametrizations was constructed by constraining $a_{\rm surf}$, evaluated within the modified Thomas--Fermi approximation, to a prescribed sequence of values, while the remaining EDF parameters were simultaneously readjusted through the global optimization procedure. Calculations with the resulting SLy5sX parametrizations revealed a clear, approximately linear correlation between $a_{\rm surf}$ and the fission barrier heights for \nuc{Hg}{180}, \nuc{Ra}{266}, and \nuc{Pu}{240}. These results established a direct empirical connection between the surface properties encoded in the EDF and its microscopic predictions for fissioning nuclei. Nevertheless, a comprehensive statistical analysis of the dependence of fission barrier heights on the underlying EDF parameters, together with their correlation with $a_{\rm surf}$, is still missing. Such an analysis is essential for quantifying the parametric uncertainty of predicted fission barrier heights and for identifying the microscopic couplings that primarily drive the observed correlation.

In this work, we investigate the microscopic origin of the correlation between the surface energy coefficient and fission barrier heights within covariant density functional theory (CDFT), employing the point-coupling functional PC-PK1~\cite{Zhao:2010PRC}. Rather than constructing a family of EDF parametrizations by directly constraining $a_{\rm surf}$, we vary the underlying EDF coupling constants within their allowed ranges and determine both the surface energy coefficients and fission observables self-consistently. This strategy allows one to trace how variations in the microscopic interaction propagate simultaneously to the surface properties of the functional and to the predicted fission barriers. Specifically, we determine the isoscalar and isovector surface energy coefficients, $a_{\rm surf}$ and $a_{\rm ssym}$, from SINM within the relativistic Thomas--Fermi approximation, and calculate the fission properties of \nuc{Pu}{240} and \nuc{U}{236}, including the fission barrier heights, the fissility parameter $x$, and the Coulomb energy $E_{\rm Coul}$. We then employ Gaussian-process emulators to construct the posterior distributions of these quantities. Two independent sets of constraints are considered: one based on the ground state properties of \nuc{Pu}{240} and \nuc{U}{236}, and the other on the saturation properties of INM. Using the resulting posterior distributions, we perform correlation and sensitivity analyses to quantify the relationships between the surface energy coefficients and fission observables and, more importantly, to identify the microscopic EDF couplings responsible for these correlations.

The remainder of this paper is organized as follows. Sec.~\ref{sec:methods} introduces the theoretical framework, including the calculation of potential-energy surfaces within CDFT, the properties of infinite nuclear matter, and the extraction of the surface energy coefficients $a_{\rm surf}$ and $a_{\rm ssym}$ from SINM. Sec.~\ref{sc:results} describes the numerical methods, including the construction and validation of the Gaussian-process emulators and the Bayesian framework employed in the subsequent analysis. The section 
also presents the results of the correlation and sensitivity analyses. Finally, Sec.~\ref{sec:summary} summarizes the main findings and discusses future perspectives.

\section{Theoretical framework}
\label{sec:methods}

\subsection{Fission barrier from CDFT}
In the relativistic mean-field (RMF) theory, the nuclear ground state is described by a Slater determinant of single-particle wave functions  $\psi_k(\bm{r})$. Accordingly, the energy of the nucleus is expressed as an energy density functional (EDF) of various densities and currents~\cite{Burvenich:2002PRC,Zhao:2010PRC}:
\begin{equation}
E_{\mathrm{DF}}\left[\mathcal T, \rho, \nabla\rho\right]=\int d^3r \left(\mathcal T(\vec{r})+\mathcal{E}^{\text {int}}(\vec{r})+\mathcal{E}^{\text {em}}(\vec{r})\right).
\label{e1}
\end{equation}
The first term is the kinetic energy:
\begin{equation}
\mathcal T(\vec{r})=\sum_kv_k^2\psi_{k}^{\dagger}(\vec{r})(\boldsymbol{\alpha} \cdot \vec{p} +\beta m-m) \psi_{k}(\vec{r}),
\end{equation}
where $  m  $  is the nucleon bare mass and $\boldsymbol{\alpha}$, $\beta$ are the Dirac matrices. The pairing effect is taken into account in the BCS approximation, where $v_k^2$ represents the occupation probability of the $k$-th single-particle state.  

In the present work, we consider the time-reversal invariant systems, for which the spatial components of the vector currents vanish. Under this condition, the nucleon-nucleon interaction part $\mathcal{E}^{\text {int }}(\vec{r})$ in Eq.~(\ref{e1}) can be expressed by:
\begin{equation}
\begin{aligned}
\mathcal{E}^{\text {int }}(\vec{r})= & \frac{\alpha_{S}}{2} \rho_{S}^{2}+\frac{\beta_{S}}{3} \rho_{S}^{3}+\frac{\gamma_{S}}{4} \rho_{S}^{4}+\frac{\delta_{S}}{2} \rho_{S} \Delta \rho_{S} \\
& +\frac{\alpha_{V}}{2} \rho_V^2+\frac{\gamma_{V}}{4}\rho_V^4+\frac{\delta_{V}}{2} \rho_V \Delta \rho_V \\
& +\frac{\alpha_{T V}}{2} \rho_{T V}^2+\frac{\delta_{T V}}{2} \rho_{T V} \cdot \Delta\rho_{T V},
\end{aligned}
\label{e2}
\end{equation}
and the electromagnetic $\mathcal{E}^{\text {em }}(\boldsymbol{r})$ interaction is:
\begin{equation}
\label{eq:Coul}
 \mathcal{E}^{\rm em}(\vec{r})
 =\frac{e}{2} A_0(\vec{r}) \rho_{V}^{(p)}(\vec{r}),
\end{equation}
where $e$ is the charge of bare proton and $A_0(\vec{r})$ represents the Coulomb potential. Owing to the long-range nature of the Coulomb interaction, the Coulomb potential is calculated using the Green's function method~\cite{GAMBHIR1990132}. The densities and the time-like components of the currents in the EDF are given as bilinear combinations of the single-particle Dirac
wave functions $\psi_k(\vec{r})$:
\begin{subequations}
\begin{align}
 \rho_S(\vec{r})          &=\sum_{k }v_k^2\bar\psi_k(\vec{r})\psi_k(\vec{r}),\\
  \rho_{V}(\vec{r})    &=\sum_{k }v_k^2\bar\psi_k(\vec{r})\gamma^{0}\psi_k(\vec{r}),\\
  \label{E13c}
  \rho_{T V}(\vec{r}) &=\sum_{k }v_k^2\bar\psi_k(\vec{r})\tau_3\,\gamma^{0}\psi_k(\vec{r}).
\end{align}
\end{subequations}
The index $k$ runs over all positive-energy single-particle states under the no-sea approximation~\cite{Ring:1996}. Minimization of the EDF with respect to $\psi^\dagger_k  $ leads to the Dirac equation:
 \begin{align}
     \label{eq:DiracEq}
\left[ \boldsymbol{\alpha} \cdot \vec{p} + \beta\left(m + S\right) + V\right] \psi_k = \epsilon_k \psi_k ,
 \end{align}
 where  
 \begin{subequations}
 \begin{align}
  S           &=\alpha_S\rho_S+\beta_S\rho^2_S+\gamma_S\rho^3_S+\delta_S\triangle\rho_S,\\
  V   &=\alpha_V\rho_V +\gamma_V (\rho_V)^3
                       +\delta_V\triangle \rho_V + e A^0 \notag\\
  &\quad\;+ \alpha_{TV}\rho_{T V}\cdot\tau_3+\delta_{TV}\triangle\rho_{T V}\cdot\tau_3.
 \end{align}
 \end{subequations}
The pairing correlation between nucleons is considered using a $\delta$ force, in which case, the pairing energy reads~\cite{Bender:2000xk,Yao:2009PRC},
\begin{equation}
E_{\text{pair}}^\tau\left[\kappa,\kappa^*\right]=-\frac{V_\tau}{4} \int d^3\vec{r} \kappa_\tau^*(\vec{r})\kappa_\tau(\vec{r}),
\end{equation}
with the pairing tensor given by
\begin{align}
\kappa(\vec{r})&=-2\sum_{k}f_ku_{k} v_{k} |\psi_k(\vec{r})|^2.
\end{align}
Here, $u_k^2=1-v_k^2$ and $f_k$ is the smooth cut-off function used to simulate the finite range effect of the pairing interaction~\cite{Krieger:1990pks}.
$\tau=n,~p$ distinguishes neutrons and protons. 

In order to generate the potential energy surface as a function of the deformation $\beta_\lambda^c$, we add a quadratic constraint term in the variational calculation,
\begin{align}
V_{\text{ext}}=\sum_{\lambda}C_{\lambda}(\beta_\lambda-\beta_\lambda^c)^2
\end{align}
where $\beta_\lambda$ is defined as
 \begin{align}
 \beta_\lambda = \frac{4\pi}{3AR^\lambda} \braket{\Phi(\beta_\lambda)|\hat{Q}_{\lambda 0}|\Phi(\beta_\lambda)},
 \end{align}
with $  R = 1.2A^{1/3}  $ fm and the multipole moment operators $  \hat{Q}_{\lambda 0} \equiv r^\lambda Y_{\lambda 0}  $. 
Here, $\ket{\Phi(\beta_\lambda)}$ denotes the mean-field wave function of atomic nucleus so obtained. 

Moreover, the microscopic center-of-mass and rotational corrections are included in the total energy as~\cite{Warda2002,Bender2004}
\bsub
\begin{align}
E_{\text{cm}}^{\text{mic}}
&=
-\frac{1}{2Am}
\braket{\Phi(\beta_\lambda)|\bm{\hat{P}}_{\text{cm}}^{2}|\Phi(\beta_\lambda)},\\
E_{\rm rot}
&=
-\frac{1}{2\mathcal{J}}
\braket{\Phi(\beta_\lambda)|\hat{J}^{2}|\Phi(\beta_\lambda)} ,
\end{align}
\esub
where $\bm{\hat{P}}_{\rm cm}$ denotes the total center-of-mass momentum operator, and the rotational moment of inertia $\mathcal{J}(\beta_\lambda)$ is evaluated using the Belyaev formula~\cite{osti_4262925}. Including these two corrections, the total energy at a given deformation $\beta_\lambda$ is given by
\begin{align}
E_{\rm tot}(\beta_\lambda)
=
E_{\mathrm{DF}}
+\sum_{\tau=n,p}E_{\text{pair}}^\tau
+E_{\text{cm}}^{\text{mic}}
+E_{\rm rot}.
\end{align}

\subsection{Surface energy and fission fissility from liquid drop model}
In the liquid drop model (LDM), the binding energy of an atomic nucleus ($^AZ$) with spherical shape is expanded as~\cite{Myers1966}:
\begin{align}
    E_{\rm LDM}=&\;
\left(a_{\rm vol}+a_{\rm sym}I^2\right)A +\left(a_{\rm surf}+a_{\rm ssym}I^2\right)A^{2/3} \nonumber\\
&+\frac{3e^2}{5r_0}\frac{Z^2}{A^{1/3}}
-\frac{3e^2}{4r_0}
\left(\frac{3}{2\pi}\right)^{2/3}
\frac{Z^{4/3}}{A^{1/3}}, 
\label{eq:LDM}
\end{align}
which includes the surface energy 
\begin{equation}
\label{eq:surf_energy}
E_{\rm surf}
=
\left(a_{\rm surf}+a_{\rm ssym}I^2\right)A^{2/3}.
\end{equation}
Here, $I=(N-Z)/(N+Z)$ is the isospin asymmetry and the parameter $r_0$ is given by $ r_0=[3/(4\pi\rho_0)]^{1/3}$ at the saturation density $\rho_0$ of infinite nuclear matter. The bulk parameter $a_{\rm vol}$ and $a_{\rm sym}$ in LDM can be extracted from the saturation properties of infinite nuclear matter, while $a_{\rm surf}$ and $a_{\rm ssym}$ are related to the properties of semi-infinite nuclear matter which will be shown later. 

For small quadrupole deformations $\beta_2$ around the spherical configuration, the surface and the Coulomb energies can be expanded to the second order in $\beta_2$ as~\cite{Bohr1939}
\begin{subequations}
\begin{align}
E_{\rm surf}(\beta_2)
&\approx E_{\rm surf}\left(1+\frac{1}{2\pi}\beta_2^2\right),\\
E_{\rm Coul}(\beta_2)
&\approx E_{\rm Coul}\left(1-\frac{1}{4\pi}\beta_2^2\right),
\end{align}
\end{subequations}
where $E_{\rm surf}$ and $E_{\rm Coul}$ denote the surface and the Coulomb energies for the spherical configuration, respectively. The corresponding change in the liquid drop energy is therefore
\begin{equation}
\delta E_{\rm LDM}(\beta_2)
=\frac{1}{2\pi}E_{\rm surf}(1-x)\beta_2^2,
\end{equation}
with the fissility parameter defined as~\cite{Hasse1988,Denschlag2011}, 
\begin{equation}
\label{eq:x}
x\equiv\frac{E_{\rm Coul}}{2E_{\rm surf}}\simeq 0.02 Z^2/A.
\end{equation}
For $x<1$, the surface energy dominates over the destabilizing Coulomb contribution, and the spherical configuration is stable against small quadrupole deformations. In contrast, for $x\geq1$, the Coulomb repulsion can no longer be compensated by the surface tension, rendering the spherical configuration unstable against infinitesimal deformations toward fission. For superheavy nuclei with $Z\gtrsim104$, the fissility parameter \(x\) given by LDM approaches unity, and the fission barrier almost disappears. In this case, the fission barrier arises from quantal shell effects. In the present calculations, $E_{\rm Coul}$ is obtained from Eq.~(\ref{eq:Coul}) under the spherical constraint ($\beta_2=0$), rather than from the semiclassical liquid drop expression. 

\subsection{Nuclear matter properties from CDFT}
Before presenting the extraction of the surface energy coefficient from semi-infinite nuclear matter, we briefly summarize the relevant properties of uniform nuclear matter that are used throughout this work.
In uniform nuclear matter with asymmetry $I$, the single-particle state is labeled with the momentum $\vec{p}=\hbar\vec{k}$ and the spin projection quantum number $\lambda$. The corresponding single-particle wave function is simplified into $u_\tau(\vec{k},\lambda)$, which is determined by the following Dirac equation 
\begin{equation}
\label{eq:nmdeq}
    \left(\bm{\alpha}\cdot\vec{k} + \beta m^*_\tau
     +\Sigma_{\tau}^{0}\right)u_\tau(\vec{k},\lambda) 
    = E_\tau(\vec{k}) u_\tau(\vec{k},\lambda),
\end{equation}
 where $m^*_\tau = m_\tau +  \Sigma_{S}$ is the Dirac mass. 
The nucleon self-energies  are determined by the scalar and the vector densities as 
\begin{subequations}\label{eq:nmsnp}
\begin{align}
    &\Sigma_{S} = \alpha_S\rho_S+\beta_S\rho_S^2
    +\gamma_S \rho_S^3,
        \\
    &\Sigma_{\tau}^{0}=\alpha_V\rho_V +\gamma_V\rho^3_V
    + \alpha_{TV}\tau_3\rho_{TV},
\end{align}
\end{subequations}
where $\rho_S=\rho_S^{(n)}+\rho_S^{(p)}$ and $\rho_V=\rho_V^{(n)}+\rho_V^{(p)}$. The scalar density $\rho_S^{(\tau)}$ and the vector density $\rho_V^{(\tau)}$ are related by the corresponding Fermi momenta $k_{\tau F}$ as 
\begin{subequations}\label{eq:rhoS_def}
\begin{align}
      \rho_S^{(\tau)}&=\frac{2}{(2\pi)^{3}}\!\int_{0}^{k_{\tau F}}\!
  \frac{m^{*}_\tau}{E^{*}_{\tau}(\vec{k})}\,d^{3}k ,\\
   \rho^{(\tau)}_{V} &=\sum_\lambda\int^{k_{\tau F}}
        \frac{d^3k}{(2\pi)^3}=\frac{k_{\tau F}^3}
        {3\pi^2},
\end{align}
\end{subequations}
with $E^{*}_{\tau}(\vec{k})=\sqrt{\vec{k}^2+m^{*\,2}_\tau}$. The corresponding energy density and the pressure of nuclear matter at the zero temperature are given by~\cite{Dutra2014,Zhang:2024_Article}
\begin{subequations}\label{eq:eps}
\begin{align}
\mathcal{E}(\rho_V,I)
=&\sum_\tau\left[\frac{3}{4}\left(
        \rho^{(\tau)}_{V} E_{\tau F}^*
        -\rho^{(\tau)}_{S}m^*_\tau\right)
        +\rho^{(\tau)}_{V}m_\tau\right]
               \notag\\
+\frac{\alpha_S}{2}&\rho_S^2
+\frac{\beta_S}{3}\rho_S^3
+\frac{\gamma_S}{4}\rho_S^4
+\frac{\alpha_V}{2}\rho_V^2
+\frac{\gamma_V}{4}\rho_V^4
+\frac{\alpha_{TV}}{2}\rho_{TV}^2,
\label{eq:energy}
\\
p(\rho_V,I)
=&\sum_\tau\left[\frac{1}{4}\left(\rho^{(\tau)}_{V}E_{\tau F}^*
    -\rho^{(\tau)}_{ S}m_\tau^*\right)+\Sigma_{\tau }^0\rho^{(\tau)}_{V} 
    +\Sigma_{S}
\rho^{(\tau)}_{S}\right]
\notag\\
-\frac{\alpha_S}{2}&\rho_S^2
-\frac{\beta_S}{3}\rho_S^3
-\frac{\gamma_S}{4}\rho_S^4
-\frac{\alpha_V}{2}\rho_V^2
-\frac{\gamma_V}{4}\rho_V^4
-\frac{\alpha_{TV}}{2}\rho_{TV}^2,
\label{eq:pressure}
\end{align}
\end{subequations}
with the isovector density $\rho_{TV}=\rho_V^{(n)}-\rho_V^{(p)}$.
The binding energy per nucleon $E/A$ and incompressibility $K_{\infty}$ are defined as 
 \begin{equation}
 E/A \equiv \mathcal{E}/\rho_V -m,\qquad K_{\infty} \equiv 9\frac{\partial p}{\partial \rho_V}.
 \end{equation}
 Moreover, the symmetry energy $E_\mathrm{sym}$ and its slope parameter $L$ are calculated by
    \begin{equation}\label{eq:nmesym}
        E_\mathrm{sym} \equiv \frac{1}{2}\left.\frac{\partial^2 (E/A)}
        {\partial I^2}\right|_{I=0},\qquad 
        L \equiv 3\rho_V
        \frac{\partial E_\mathrm{sym}}{\partial \rho_V}.
    \end{equation}
In this work, we take the bare nucleon mass to be the same for protons and neutrons, so that the effective mass $m^*$ is independent of isospin. We further define the dimensionless effective mass as $m^*/m$.
\FloatBarrier
\subsection{Surface energy coefficient from semi-infinite nuclear matter}
\label{sc:SINM}
Semi-infinite nuclear matter (SINM) is an extension of infinite nuclear matter with an introduction of a planar surface separating uniform nuclear matter from the vacuum.  The matter remains translationally invariant in the directions parallel to the surface, and the density varies only along the direction perpendicular to it. Using the relativistic Thomas--Fermi approximation~\cite{CENTELLES1998193}, the  surface tension of SINM with 
asymmetry $I$ is expressed by 
\begin{equation}
\sigma(I)=2\!\int_{0}^{\rho_0}\!
  \sqrt{C(\rho_V,I)\,w(\rho_V,I)}\;d\rho_V\; ,
  \label{eq:sigma1}
\end{equation}
which contains the excess energy density
\begin{equation}
  w(\rho_V,I)\;\equiv\;\mathcal{E}(\rho_V,I)-\mu_n\rho_V^{(n)}-\mu_p\rho_V^{(p)},
  \label{eq:w}
\end{equation}
with the chemical potential of the bulk nuclear matter $\mu_\tau$ at the saturation point $\rho_0$
\begin{align}
\label{eq:chemical_pot}
    \mu_\tau
=
\left.
\frac{\partial\mathcal E}{\partial\rho_V^{(\tau)}}
\right|_{\rho_0},
\end{align}
and the stiffness function
\begin{equation}
  C(\rho_V,I)=-\frac{1}{2}\!\left[
  \delta_V
  +\delta_S\!\left(\frac{d\rho_S}{d\rho_V}\right)^{\!2}
  +\delta_{TV}\,I^{2}
  \right] .
  \label{eq:C}
\end{equation}
Here, within the local-density approximation, the energy density $\mathcal{E}(\rho_V,I)$ is evaluated using the same energy density functional as in Eq.~(\ref{eq:energy}), while the scalar density $\rho_S$ is determined self-consistently for a given vector density $\rho_V$ from the relation in Eq.~(\ref{eq:rhoS_def}). More details are provided in Appendix.~\ref{app:sinm}.

The surface energy coefficient $\asurf(I)$ in the leptodermous expansion is given by
\begin{equation}
\asurf(I)=4\pi r_0^{2}(I) \sigma(I),
\label{eq:asurfI}
\end{equation}
whereas the isoscalar surface energy coefficient is obtained in the symmetric-matter limit $I=0$.
Expanding $\asurf(I)$ to second order in the isospin asymmetry $I$ gives
\begin{equation}
\asurf(I)
=\asurf+\assym I^{2}+\mathcal{O}(I^{4}),
\label{eq:assym}
\end{equation}
where $\assym$ is the surface symmetry coefficient that corresponds to the coefficient of the $I^2$ term. In practice, we evaluate Eq.~\eqref{eq:asurfI} over the interval $I\in[0,0.25]$ in steps of $0.05$ and determine $\assym$ from a least-squares fit of $\asurf(I)$ as a function of $I^2$.

\section{Results and discussion}
\label{sc:results}

\subsection{Numerical details}
\label{sc:nd}

The present study focuses on two representative actinide nuclei \nuc{Pu}{240} and \nuc{U}{236}. For simplicity, they are assumed to be axially  deformed. Therefore, it is convenient to expand the large and small components of the single particle wave functions in Eq.~(\ref{eq:DiracEq}) on the axially deformed harmonic oscillator (HO) basis, where the potential for the HO basis reads
\begin{equation}
V_{\text{osc}}(z, r_\perp) = \frac{1}{2} m \omega_z^2 z^2 + \frac{1}{2} m \omega_\perp^2 r_\perp^2.
\end{equation}
The oscillator frequencies $\omega_\perp$ and $\omega_z$ are chosen to be dependent on the deformation parameter $\beta_{\rm HO}$,
\begin{subequations}
\begin{align}
\label{eq:HO_deform}
\omega_z &= \omega_0 \exp\left({-\sqrt{\frac{5}{4\pi}} \beta_{\rm HO}}\right), \\
 \omega_\perp &=  \omega_0 \exp\left({\frac{1}{2} \sqrt{\frac{5}{4\pi}} \beta_{\rm HO}}\right).
\end{align}
\end{subequations}
The reference oscillator frequency is
given by $\hbar\omega_{0}=41A^{-1/3}$ MeV. In the calculation of the potential energy curve (PEC),  $\beta_{\rm HO}$  is chosen to be $\beta_{\rm HO}=\bt$ for $\bt<0.4$ and $\beta_{\rm HO}=\bt/2$ for $\bt \ge0.4$, which improves the convergence of the iteration solution of Eq.~(\ref{eq:DiracEq}) at large deformation. The spatial integrations are performed using 32 Gauss–Hermite quadrature points in the axial direction and 32 Gauss–Laguerre quadrature points in the radial direction. Convergence tests confirm that this choice is sufficient to ensure numerical convergence. 
The number of HO major shells is taken to be $N_f=18$ in all subsequent calculations. The residual deviation from the converged ($N_f\to\infty$) result is accounted for as a systematic uncertainty in the Bayesian analysis described below.

\subsection{Potential energy curves and fission paths with PC-PK1}
\begin{figure}[htbp]
    \centering
    \includegraphics[width=\linewidth]{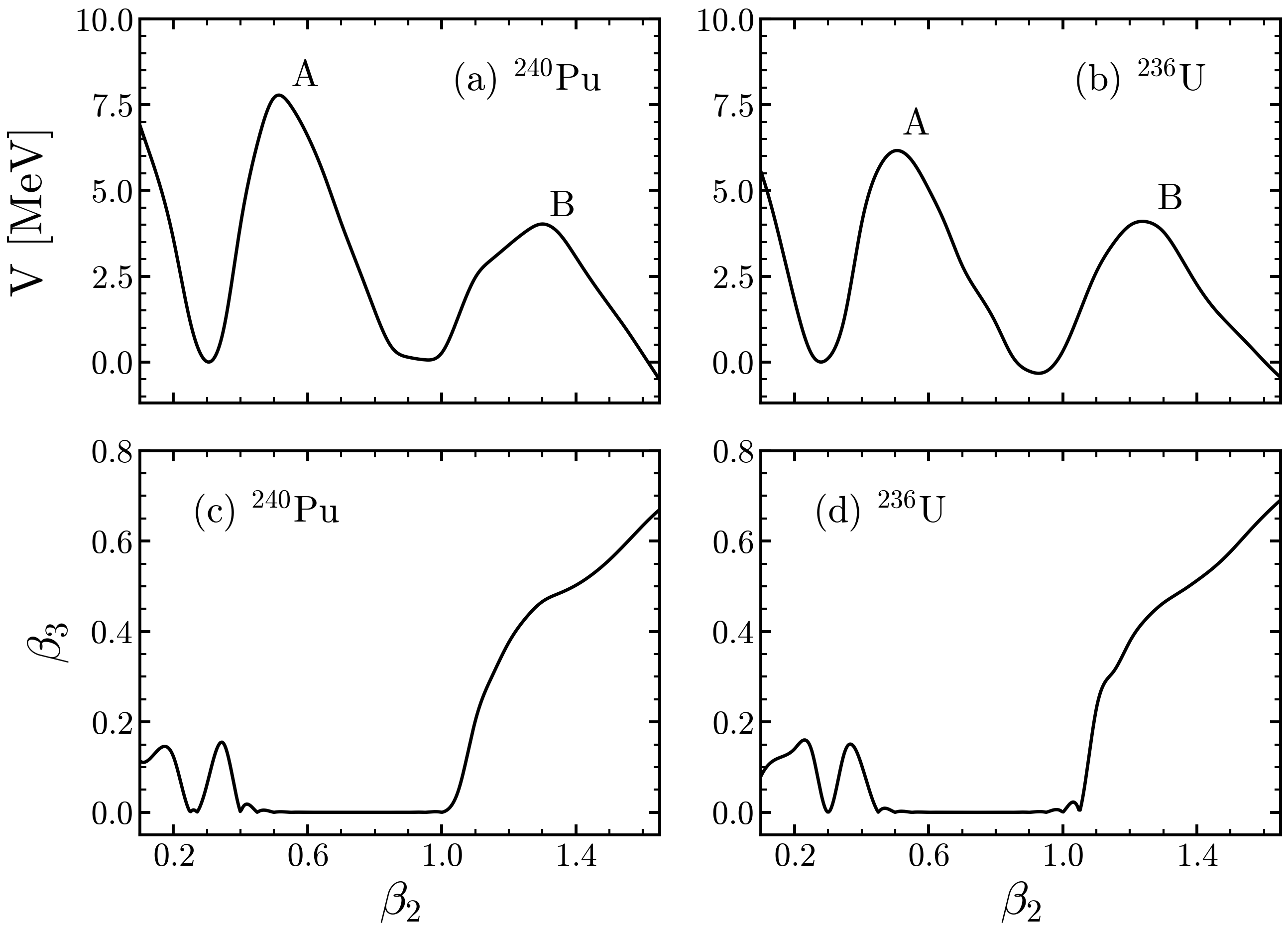}
    \caption{Panels (a) and (b) show the potential energy curves as functions of the quadrupole deformation $\beta_2$ for \nuc{Pu}{240} and \nuc{U}{236}, respectively, while the panels (c) and (d) show the corresponding fission paths. \(A\) and \(B\) denote the inner and outer fission barriers, respectively.
    These are obtained with the covariant energy functional with the PC-PK1 parameter set.     
    }
    \label{fig:PEC}
\end{figure}
Figures~\ref{fig:PEC} (a) and (b) show the potential energy curves as functions of $\bt$ for \nuc{Pu}{240} and \nuc{U}{236}, respectively, while panels (c) and (d) show the corresponding fission paths, calculated with the PC-PK1 parametrization using 18 major shells.  Here, the value of $V(\bt)$ is defined as $V(\bt) = E(\bt) - E_{\rm g.s.}$. The fission path $\beta_3^{\rm fp}(\beta_2)$ is defined by the fission valley in the $(\bt,\beta_3)$ two-dimensional potential energy surface, that is, for each $\bt$ along the fission path, the value of $\beta_3$ is chosen to minimize the energy. The inner and outer fission barriers are denoted by \(A\) and \(B\) in the figure.

In Fig.~\ref{fig:PEC} (b) for \nuc{U}{236}, the inclusion of the rotational correction lowers the energy of the superdeformed minimum ($\beta_2\sim0.95$) below that of the first minimum ($\beta_2\sim0.3$). Nevertheless, throughout this work, the ground state is identified with the first minimum, and all fission barrier heights are defined relative to it. This convention is adopted because the first minimum represents the normal-deformed ground state configuration, whereas the other minima correspond to shape-isomeric configurations along the fission path. In this work, we use $E_A$ and $E_B$ to indicate the inner and outer fission barrier height.

The reference values of the ground state and the fission properties calculated with PC-PK1 with \(N_f=18\)
are summarized in Table~\ref{tab:PC-PK1}. The table also lists the surface energy coefficients, $a_{\rm surf}$ and $a_{\rm ssym}$, calculated from the method described in Sec.~\ref{sc:SINM}.

\begin{table}[htbp]
\centering
\renewcommand{\arraystretch}{1.35}
\caption{The reference values of various quantities for \nuc{Pu}{240} and \nuc{U}{236} calculated with PC-PK1. These quantities include the ground state energy $E_{\rm g.s.}$, the root-mean-square (rms) charge radius $r_p$, the inner fission barrier height $E_A$, the outer fission barrier height $E_B$, the Coulomb energy $E_{\rm Coul}$ at the spherical shape, the fissility parameter $x$, the surface energy coefficient $a_{\rm surf}$, and the surface symmetry energy coefficient $a_{\rm ssym}$.}
\label{tab:PC-PK1}
\begin{ruledtabular}
\begin{tabular}{lll}
Quantities & \nuc{Pu}{240} & \nuc{U}{236} \\
\hline
$E_{\rm g.s.}$ (MeV)   & $-1812.382$ & $-1788.803$\\
$r_p$ (fm)           & $5.932$ & $5.879$    \\
$E_A$ (MeV)          & $7.778$    & $ 6.164$     \\
$E_B$ (MeV)          & $4.020$       & $ 4.099$   \\
$E_{\rm Coul}$ (MeV)    & $1017.351$  & $981.192$  \\
$x$                    & $0.755$     & $0.738$   \\
$a_{\rm surf}$ (MeV) & $18.849$   & $18.849$      \\
$a_{\rm ssym}$ (MeV) & $-29.896$   & $-29.896$ 
\end{tabular}
\end{ruledtabular}
\end{table}

\subsection{Emulator}
\label{sec:emulator}

To investigate the correlation between the fission barrier height and the surface energy coefficient, it is required to scan 
a sufficiently large ensemble of EDF parameter sets. We first generate 200 EDF parameter sets by varying the nine coupling constants, $\vec{C}=[\alpha_S,\beta_S,\gamma_S,\delta_S,\alpha_V,\gamma_V,\delta_V,\alpha_{TV},\delta_{TV}]$ around their PC-PK1 values using Latin hypercube sampling~\cite{Dutta:2020LHS}, which provides an efficient and representative sampling of a multidimensional parameter space. The sampling ranges are taken from Table II of Ref.~\cite{Zhang:2024_Article}, where they were determined from the uncorrelated tolerance corresponding to \(\chi^2 \leq \chi^2_{\rm min}+1\), following Ref.~\cite{Burvenich:2002PRC}. After obtaining the 200 EDF parameter sets, we calculate the corresponding fission properties and surface energy coefficient.

To increase the statistical precision of the analysis, Gaussian-process (GP) emulators~\cite{Rasmussen2006} are employed to enlarge the ensemble of EDF parameter sets. Here, GP emulators provide a probabilistic surrogate for the mapping between the EDF parameters and the observables, together with an estimate of the predictive uncertainty. Given the training set \(\mathcal D\), the GP provides the predictive distribution of an observable \(O\) for the parameter set \(\vec C\):
\begin{equation}
\label{eq:GP}
p(O|\mathcal{D},\vec{C}) = \mathcal{N}(\mu_{\rm GP}(\vec{C}),\;\sigma_{\rm GP}^2(\vec{C})),
\end{equation} 
where $ \mathcal{N}(\mu_{\rm GP}(\vec{C}),\,\sigma_{\rm GP}^2(\vec{C}))$ denotes the multivariate normal distribution with the mean $\mu_{\rm GP}(\vec{C})$ and the variance $\sigma_{\rm GP}^2(\vec{C})$.
Here, independent scalar GPs are employed
to derived scalar observables
($E_{\rm g.s.},r_p,E_A,E_B,E_{\rm coul},x ,\asurf,\assym$) and the nuclear matter properties. Meanwhile, instead of constructing scalar GP for each deformation grid of the full PEC $V(\bt)$ and the fission path $\beta_3^{\rm fp}(\beta_2)$, the principal component analysis (PCA)-GP~\cite{Rasmussen2006} is used to capture their principal modes.  The benchmark of the GP emulator is presented in Appendix~\ref{app:benchmark}, which shows that the emulator precision converges with 120 training sets. In the present work, we use all 200 generated samples as the training set $\mathcal{D}$.

Once the GP model is trained, the posterior predictive distribution of any quantity $O^{\rm pred}$ can be constructed within the Bayesian framework,
\begin{equation}
  \label{eq:posterior_O_Data}
  p(O^{\rm pred}|\mathcal{D},\vec{Q}^{\rm ref})
  =\int p(O^{\rm pred}|\mathcal{D},\vec{C})\;
p(\vec{C}|\mathcal{D},\vec{Q}^{\rm ref})\;d\vec{C} ,
\end{equation}
where $\vec{Q}^{\rm ref}$ denotes the reference values used to constrain the EDF parameters. Here, $p(\vec{C}|\mathcal{D},\vec{Q}^{\rm ref})$ is the parameter posterior:
\begin{align}
  \label{eq:post_parameter_data}
p(\vec{C}|\mathcal{D},\vec{Q}^{\rm ref})
=\frac{\pi(\vec{C})\,p(\vec{Q}^{\rm ref}|\mathcal{D},\vec{C})}
    {p(\mathcal{D},\vec{Q}^{\rm ref})} ,
\end{align}
with the normalization constant $p(\mathcal{D},\vec{Q}^{\rm ref})$ and the prior $\pi(\vec{C})$. $ p(O^{\rm pred}|\mathcal{D},\vec{C})$ is the predictive probability of the quantity $O^{\rm pred}$ in the converged ($N_f\to\infty$) limit, which is expressed by
\begin{align}
\label{eq:predictive_distribution}
  p(O^{\rm pred}|\mathcal{D},\vec{C})
  =\int p(O^{\rm pred}|O)\;  p(O|\mathcal{D},\vec{C})\;dO ,
\end{align}
where $O$ denotes the corresponding unknown quantity calculated with 
a finite value of $N_f$, that is, $N_f=18$ in the present case. Here, the predictive probability  $p(O|\mathcal{D},\vec{C})$ 
in Eq.~(\ref{eq:predictive_distribution}) is provided by the GP emulator as formalized in Eq.~(\ref{eq:GP}), and the condition probability is given by
\begin{equation}
    p(O^{\rm pred}|O)
=\mathcal{N}\big(O,\;\sigma_{\rm sys}^2\big),
\end{equation} 
which contains the systematic error $\sigma_{\rm sys}$ induced by the basis truncation. The value of $\sigma_{\rm sys}$ for $O^{\rm pred}\in\{V(\bt), \allowbreak\beta_3^{\rm fp}(\beta_2), \allowbreak E_{\rm g.s.}, \allowbreak r_p,E_A,E_B,E_{\rm coul},x\}$ is estimated from the difference between the results computed with $N_f=18$ and $N_f=20$. This choice is supported by Ref.~\cite{Lu2014}, which demonstrated that calculations with $N_f=20$ are already converged.

For the parameter posterior in Eq.~(\ref{eq:post_parameter_data}), we adopt a uniform prior over the sampling range of each coupling constant. Here, the marginal likelihood $p(\vec{Q}^{\rm ref}|\mathcal{D},\vec{C})$ is calculated as, 
\begin{align}
  \label{eq:likelihood}
  p(\vec{Q}^{\rm ref}|\mathcal{D},\vec{C})
  =\int p(\vec{Q}^{\rm ref}|\vec{Q})\;p(\vec{Q}|\mathcal{D},\vec{C})\;d\vec{Q},
\end{align}
where $\vec{Q}$ denotes the unknown values of the corresponding quantities at the parameter set $\vec{C}$ with $N_f=18$. Here, $p(\vec{Q}|\mathcal{D},\vec{C})$ is also provided by the GP emulator. The condition probability $p(\vec{Q}^{\rm ref}|\vec{Q})$ is approximated as
\begin{equation}
\label{eq:condition_probability}
  p(\vec{Q}^{\rm ref}|\vec{Q})
  =\mathcal{N}\big(\vec{Q},\;\Sigma_{\rm tol}\big) ,
\end{equation}
where the covariance matrix $\Sigma_{\rm tol}$ quantifies the tolerances of the reference values $\vec{Q}^{\rm ref}$. In the present work, $\Sigma_{\rm tol}$ is assumed to be diagonal, with each diagonal element $\sigma^2_{\rm tol}$ corresponding to the square of the tolerance for the respective quantity. 

To compare the effects of imposing constraints on ground state properties and nuclear matter saturation properties, we consider two sets of reference data, $\vec{Q}^{\rm ref}_1=\{E_{\rm g.s.}, r_p\}$ and $\vec{Q}^{\rm ref}_2=\{\rho_0, E/A, E_{\rm sym},L, K_{\infty},m^*/m\}$. 
For \(\vec{Q}^{\rm ref}_1\), the reference values of the ground state binding energies and charge radii are listed in Table~\ref{tab:PC-PK1}.  For the tolerance values, we use \(\sigma_{\rm tol}=1\) MeV and \(0.02\) fm for $E_{\rm g.s.}$ and $r_p$, respectively. 
The constraint from $\vec{Q}^{\rm ref}_2$ is found to be weaker than that from $\vec{Q}^{\rm ref}_1$ for all the quantities considered, and both the reference sets lead to consistent statistical conclusions. Therefore, we only present the results obtained with $\vec{Q}^{\rm ref}_1$ in the main text,  while the corresponding results for $\vec{Q}^{\rm ref}_2$ are provided in the Supplemental Material.

\subsection{Uncertainty quantification and correlation analysis}
\label{sc:rd}

\begin{figure}[htbp]
\includegraphics[width=1.02\columnwidth]{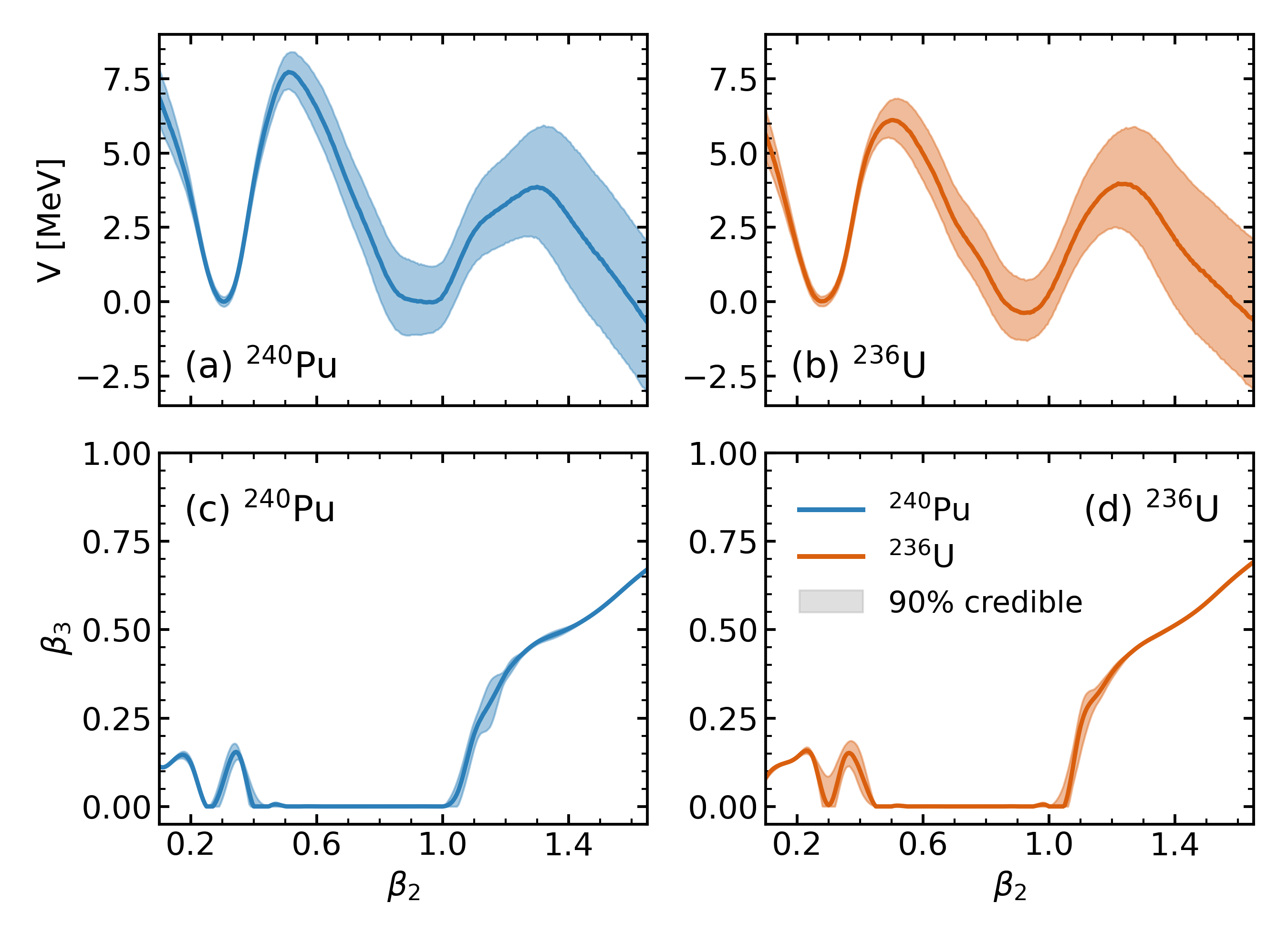}
\caption{
Similar to Fig.~\ref{fig:PEC}, but for a collection of the results with 6000 different parameter sets of the energy density functional. 
See the text for details. 
In all the panels, the solid curve is the median value of the posterior distribution and the shaded band is the 90\% credible interval.}
\label{fig:bands}
\end{figure}

Using the trained PCA-GP emulator, we predict the PECs and corresponding fission paths for 6000 EDF parameter sets drawn from the posterior distribution. Figures~\ref{fig:bands} (a) and (b) show the median and 90\% credible bands of the PECs from the predictions for \nuc{Pu}{240} and \nuc{U}{236}, respectively. Figs.~\ref{fig:bands} (c) and (d) show the corresponding fission paths. From the figure, the fission paths of the two nuclei exhibit similar behaviors. Since the energy is normalized to zero at the first minimum ($\beta_2\sim0.3$), the small uncertainty around this minimum indicates that the corresponding deformation $\beta_2$ is highly stable against different EDFs. For deformations beyond the first minimum, the uncertainty generally increases with increasing $\beta_2$. In addition, the uncertainty around the second fission barrier is larger than that around the first barrier, indicating that the second barrier is more sensitive to the variations in the EDF parameters. The fission paths, in contrast, exhibit rather narrow credible bands over the entire deformation range, indicating that the equilibrium octupole deformation along the fission valley is insensitive to the EDF parameter set.

\begin{table}[htbp]
\centering
\renewcommand{\arraystretch}{1.3}
\caption{
The mean ($\mu$) and the standard deviation ($\sigma$) values of the prior and the posterior distributions 
calculated from the GP emulators for different quantities of $^{240}$Pu and $^{236}$U. 
These are presented in the notation of $\mu\,(\sigma)$. 
See Table \ref{tab:PC-PK1} for the meaning of each quantity. 
The values are estimated using importance sampling with 600,000 samples.}
\label{tab:summary}
\begin{ruledtabular}
\begin{tabular}{lll}
Quantities & Prior & Posterior \\
\hline
\multicolumn{3}{c}{$^{240}$Pu} \\
\hline
$E_{\rm g.s.}$ (MeV)   & $-1811.462\,(55.214)$  & $-1812.368\,(1.010)$  \\
$r_p$ (fm)           & $5.932\,(0.045)$       & $5.932\,(0.014)$      \\
$E_A$ (MeV)          & $7.700\,(0.592)$       & $7.755\,(0.355)$      \\
$E_B$ (MeV)          & $4.014\,(1.203)$       & $4.029\,(1.121)$      \\
$E_{\rm Coul}$ (MeV)    & $1017.571\,(7.981)$    & $1017.454\,(2.638)$   \\
$x$                    & $0.755\,(0.015)$      & $0.755\,(0.012)$ \\
$a_{\rm surf}$ (MeV) & $18.860\,(0.510)$      & $18.852\,(0.303)$     \\
$a_{\rm ssym}$ (MeV) & $-29.949\,(2.026)$     & $-29.888\,(1.735)$    \\
\hline
\multicolumn{3}{c}{$^{236}$U} \\
\hline
$E_{\rm g.s.}$ (MeV)   & $-1788.033\,(54.186)$  & $-1788.797\,(1.011)$  \\
$r_p$ (fm)           & $5.883\,(0.044)$       & $5.880\,(0.015)$      \\
$E_A$ (MeV)          & $6.139\,(0.489)$       & $6.163\,(0.380)$      \\
$E_B$ (MeV)          & $4.167\,(1.125)$       & $4.137\,(1.036)$      \\
$E_{\rm Coul}$ (MeV)    & $981.383\,(7.783)$     & $981.577\,(2.545)$    \\
$x$                    & $0.739\,(0.015)$      & $0.738\,(0.012)$ \\
$a_{\rm surf}$ (MeV) & $18.860\,(0.510)$      & $18.866\,(0.304)$     \\
$a_{\rm ssym}$ (MeV) & $-29.949\,(2.026)$     & $-29.963\,(1.733)$    \\
\end{tabular}
\end{ruledtabular}
\end{table}
 
Table~\ref{tab:summary} compares the mean $\mu$ and standard deviation $\sigma$ of the prior and posterior distributions of selected quantities for $^{240}$Pu and $^{236}$U. The values of these quantities are estimated using importance sampling with 600,000 samples. The most significant change from prior to posterior is seen in the ground state binding energy $E_{\rm gs}$ and the rms charge radius $r_p$. This is expected since the likelihood is constructed directly from the constraint of $E_{\rm gs}$ and $r_p$. The Coulomb energy $E_{\rm C}$ also shows a notable reduction in their standard deviations, reflecting that $E_{\rm C}$ is highly correlated with the properties of the ground state through the EDF parameters. 
In contrast, the fission barrier heights $E_A$ and $E_B$, and the fissility parameter $x$ for both nuclei show only minor changes between the prior and posterior. This indicates that constraining the ground state properties  provides limited additional information on the fission-related quantities. Since the fission barriers and the fissility parameter are closely related to the spontaneous fission half-lives, this suggests that incorporating experimental half-life data into the likelihood could more effectively constrain the EDF parameters relevant to fission. 

For both nuclei, the posterior distributions show that the relative uncertainties ($\sigma/|\mu|$) of the fission properties are substantially larger than those of the ground state properties. In particular, the height of the outer barrier $E_B$ has a relative uncertainty of about 25\%, significantly larger than those of the other fission properties, which are below 6\%. This indicates that $E_B$ is particularly sensitive to variations in the EDF parameters. Table~\ref{tab:summary} also provides a statistical uncertainty for the observables of $^{240}$Pu and $^{236}$U, together with $a_{\rm surf}$ and $a_{\rm ssym}$, within the PC-PK1 parametrization.

\begin{figure}[htbp]
\includegraphics[width=0.99\columnwidth]{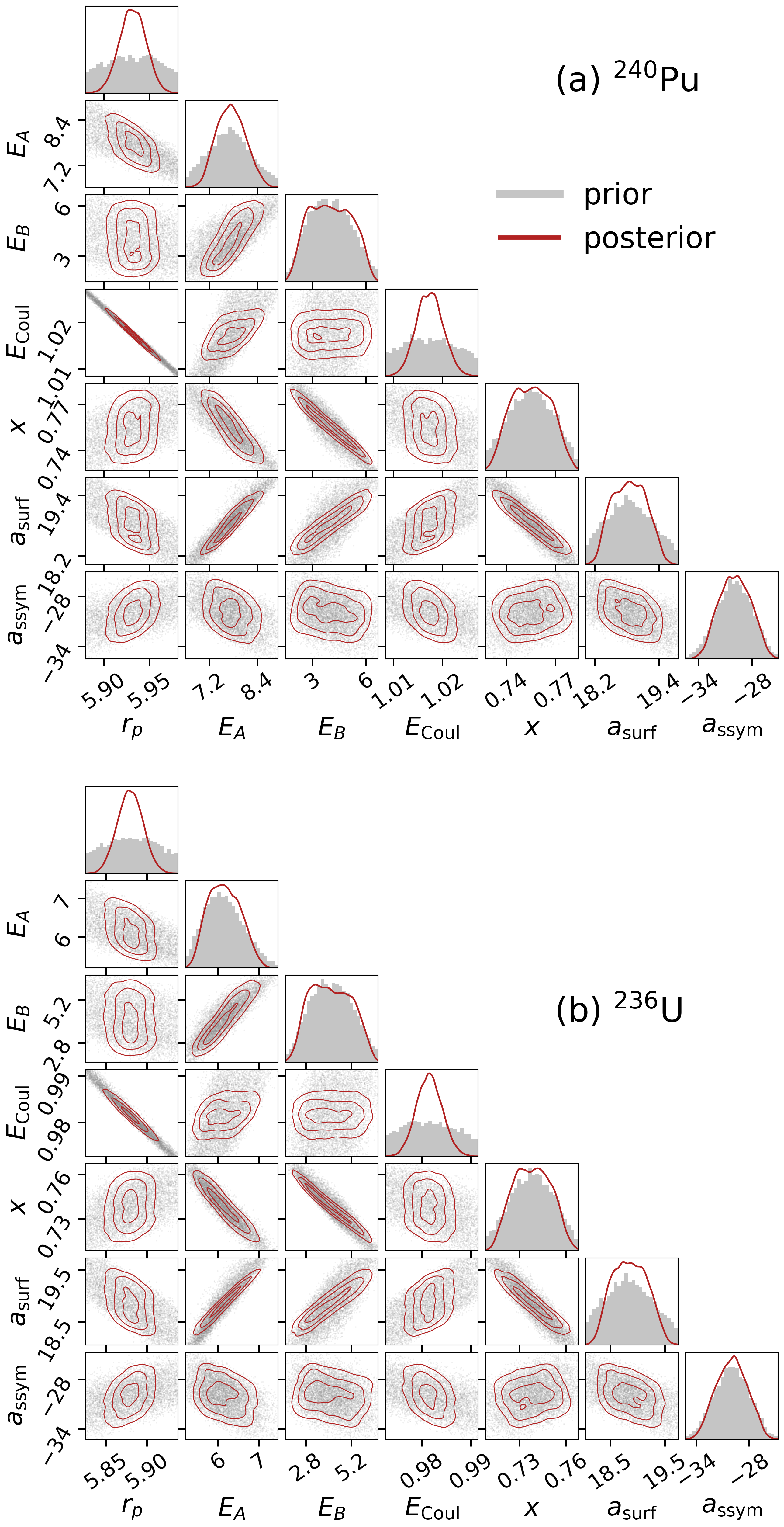}
\caption{The correlation cornerplot among different quantities. The diagonal diagrams are the histograms for the probability distributions of these quantities. The grey points and the histograms are the prior, while the red contours and the curves the posterior. The contours for the posterior represent the 39.3\%, 86.5\%, and 98.9\% highest-density regions. Panels (a) and (b) are the results for \nuc{Pu}{240} and \nuc{U}{236}, respectively. The unit of $E_{\rm Coul}$ is GeV, whereas the units of all the other quantities are the same as in Table~\ref{tab:summary}.}
\label{fig:obscorner}
\end{figure}

Figures~\ref{fig:obscorner} (a) and (b) show the pairwise correlations among the selected quantities for \nuc{Pu}{240} and \nuc{U}{236}, respectively. The cornerplot is constructed using the same 600,000 importance samples as those used for Table~\ref{tab:summary}. The gray points and histograms represent the samples from prior distribution, while the red contours and lines correspond to the posterior distribution obtained under the constraint of the ground state properties. The three contour levels enclose 39.3\%, 86.5\%, and 98.9\% of the posterior probability mass, which corresponds to the $1\sigma$, $2\sigma$, and $3\sigma$ highest-density regions of a two-dimensional Gaussian distribution.

In the prior distribution, 
one can see that deviations of the quantities from their central values tend to 
be correlated, especially  
in the region far from the center. 
This trend can be attributed to the fact that large deviations in both of the quantities in each panel 
are primarily driven by variations in $\alpha_S$ and $\alpha_V$, which provide the dominant contribution to the nuclear mean-field potential~\cite{Burvenich:2002PRC}. However, after imposing  the constraint on the ground state properties, several correlations present 
in the prior are substantially reduced in the posterior.  This highlights the importance of the statistical constraint for correlation analyses. In contrast,  several clear correlations from the posterior distributions are visible in both nuclei:
\begin{itemize}
    \item The proton rms radius $r_p$ and the Coulomb energy $E_{\rm Coul}$ show a strong anti-correlation. This can be understood from the fact that the larger proton radius corresponds to the more diffuse charge distribution, which weakens the electrostatic repulsion among protons and thus reduces the Coulomb energy.

    \item  The fissility parameter $x$ is strongly anti-correlated with the surface energy coefficient $a_{\rm surf}$, but only weakly correlated with the surface symmetry energy coefficient $a_{\rm ssym}$ and the Coulomb energy $E_{\rm Coul}$. Since $x$, defined by Eq.~\eqref{eq:x}, depends on $E_{\rm Coul}$, $a_{\rm surf}$ and $a_{\rm ssym}$, this result indicates that the variation of $x$ across the ensemble is dominated by that of $a_{\rm surf}$. We expect this dominance to be nucleus-dependent, as the contribution of \(a_{\rm ssym}\) in Eq.~\eqref{eq:surf_energy} is suppressed by the relatively small values of \(I^2\) for \nuc{Pu}{240} and \nuc{U}{236} (\(I^2\approx0.05\)).
        
    \item  $a_{\rm surf}$ is positively correlated with both the fission barrier heights $E_A$ and $E_B$. This is physically expected: the larger surface energy coefficient leads to the stiffer nucleus surface, which costs more energy to deform, leading to higher fission barriers. The correlation between $a_{\rm surf}$ and $E_A$ is noticeably stronger than that between $a_{\rm surf}$ and $E_B$. As can be seen from the fission paths in Fig.~\ref{fig:bands}, the inner barrier occurs at a deformation $(\beta_2,\beta_3)\approx (0.5,0)$ for both \nuc{Pu}{240} and \nuc{U}{236}. In this regime, the barrier height is primarily governed by the competition between the surface energy and the Coulomb energy. Consequently, $E_A$ is more tightly correlated with $a_{\rm surf}$. In contrast, the outer barrier occurs at $(\beta_2,\beta_3)\approx (1.3,0.5)$, where the nucleus is highly elongated with a neck formation. At such large deformations, the nuclear surface develops a pronounced curvature, whose energy contribution is not captured by the surface energy coefficient $a_{\rm surf}$. As a result, the outer barrier height $E_B$ exhibits a weaker correlation with $a_{\rm surf}$ than $E_A$. Since the variation of \(x\) is dominated by that of \(a_{\rm surf}\), and \(a_{\rm surf}\) is correlated with the fission barrier heights, \(x\) is consequently correlated with the fission barrier heights, as shown in the figure. This behavior is consistent with the classical liquid drop picture, in which a larger fissility indicates that the Coulomb repulsion more effectively counteracts the surface tension, thereby lowering both the fission barriers. However, unlike the case of $a_{\rm surf}$, the correlation between $x$ and $E_A$ is slightly weaker than that between $x$ and $E_B$.

    \item  $E_A$ and $E_B$ show a moderate positive correlation in the posterior distribution for both nuclei. This correlation is somewhat stronger for \nuc{U}{236} than for \nuc{Pu}{240}, indicating that the degree of correlation between the two barriers is nucleus dependent.
\end{itemize}
The proton charge radius $r_p$ and the Coulomb energy $E_{\rm Coul}$ evaluated for the spherical shape, show only a weak correlation with the fission barrier heights and the fissility parameter $x$. This confirms that the correlations between the ground state properties and the fission properties are generally weak. The surface symmetry energy coefficient $a_{\rm ssym}$ shows weak correlations with all the other quantities. This can be understood from the fact that $a_{\rm ssym}$ characterizes dependence of the surface tension on the 
asymmetry $I$, which is governed by the isovector channel of the functional.

\subsection{Sensitivity analysis}

To further investigate the mechanism behind the correlation between the different quantities, we perform a global sensitivity analysis (GSA) for the different quantities with respect to the nine parameters 
using the parameter posterior distribution $p(\vec{C}|\mathcal{D},\vec{Q}^{\rm ref})$ in Eq.~(\ref{eq:post_parameter_data}). From the parameter posterior distribution, one can compute the Pearson coefficients between the different parameter pairs. Those are shown in Fig.~S3 in Supplemental Material, which indicates that some of the parameters are correlated with each other. Because of this, we carry out the Shapley effect-based global sensitivity analysis (Shapley GSA)~\cite{Song2016}, which naturally accounts for the correlations among the posterior EDF parameters. 
The Shapley share for parameter $C_i\in \vec{C}$ is defined as the average marginal contribution of $C_i$ to the variance of the quantity $O^{\rm pred}$ over all possible subsets of parameters:
\begin{equation} \phi_i = \sum_{S\subseteq \vec{C}_{/C_i}} \frac{N_S!(N-N_S-1)!}{N!} \left[\mathcal V(S\cup{C_i})-\mathcal V(S)\right],
\end{equation}
where $N$ is the total number of parameters, $S$ is a subset of the parameters without $C_i$, $N_S$ is the number of parameters in $S$,  and $S\cup{C_i}$ is the subset formed by adding $C_i$ to $S$. $\mathcal V(S)$ is the variance of the conditional expectation of $O^{\rm pred}$ for given the parameters in $S$,
\begin{equation}
\mathcal V(S) = \mathrm{Var}\left[\mathbb{E}(O^{\rm pred}\mid S)\right],
\end{equation}
which measures how much of the total variance of $O^{\rm pred}$ can be explained by fixing the parameters in the subset $S$. The difference $\mathcal V(S\cup{C_i})-\mathcal V(S)$ measures how much additional variance is explained when $C_i$ is included in the subset. These variances can be evaluated using the parameter posterior distributions $ p(\vec{C}|\mathcal{D},\vec{Q}^{\rm ref})$ in Eq.~(\ref{eq:post_parameter_data}).

As we will show below, the fission properties are most sensitive to the parameter $\delta_V$ in EDF, 
and we would like to discuss the Shapley share particularly for this parameter. 
Notice that, for the correlated coupling constants from the parameter posterior distributions, the resulting Shapley shares reflect both the functional dependence of the observable on the parameters and the statistical dependence induced by their correlations, the latter of which may lead to an unphysical value of the Shapley shares. For instance, as shown in Fig. S4 in Supplemental Material, the Shapley share of $\delta_V$ for $E/A$ is large, despite that the derivative terms in EDF do not contribute to the energy of nuclear matter.  This issue is common to sensitivity analyses and is not specific to the Shapley GSA. To avoid this problem, we construct two additional factorized posterior distributions 
\begin{subequations}
\begin{align}
     p_{2}(\vec{C}|\mathcal{D},\vec{Q}^{\rm ref})
  &= p(\delta_V|\mathcal{D},\vec{Q}^{\rm ref})\,
    p(\vec{C}_{/\delta_V}|\mathcal{D},\vec{Q}^{\rm ref}),\\
    p_{3}(\vec{C}|\mathcal{D},\vec{Q}^{\rm ref})&=\prod_{i=1}^{9}p(C_i|\mathcal{D},\vec{Q}^{\rm ref}), 
\end{align}
\end{subequations}
which are used to evaluate the Shapley share, instead of $ p(\vec{C}|\mathcal{D},\vec{Q}^{\rm ref})$.  
Here,  $p(\delta_V|\mathcal{D},\vec{Q}^{\rm ref})$ and $p(\vec{C}_{/\delta_V}|\mathcal{D},\vec{Q}^{\rm ref})$ denote the corresponding marginal posterior distributions and $p(C_i|\mathcal{D},\vec{Q}^{\rm ref})$ is the marginal posterior distribution of each individual coupling constant. By constructing this way, $\delta_V$ is decorrelated from the remaining eight couplings in $ p_{2}$, and all parameters are decorrelated in $ p_{3}$. 
In the following, we denote the Shapley share evaluated with $p$, $p_2$, and $p_3$ as 
$\vec{\phi}(p)$, $\vec{\phi}(p_2)$, and $\vec{\phi}(p_3)$, respectively. 
A large discrepancy between $\vec{\phi}(p)$ and $\vec{\phi}(p_2)$  indicates that the correlation between $\delta_V$ and the other coupling constants has a substantial effect on its Shapley share, so that the value obtained from the full posterior $\vec{\phi}(p)$ cannot be interpreted as a genuine functional sensitivity. A large discrepancy between $\vec{\phi}(p_2)$ and $\vec{\phi}(p_3)$, on the other hand, indicates that the correlations among the remaining eight coupling constants are important, causing their Shapley effects to be shared among correlated parameters rather than assigned to any single one.
Here, the Shapley shares are computed using 600,000 importance samples.

\begin{figure}[htbp]
\includegraphics[width=\columnwidth]{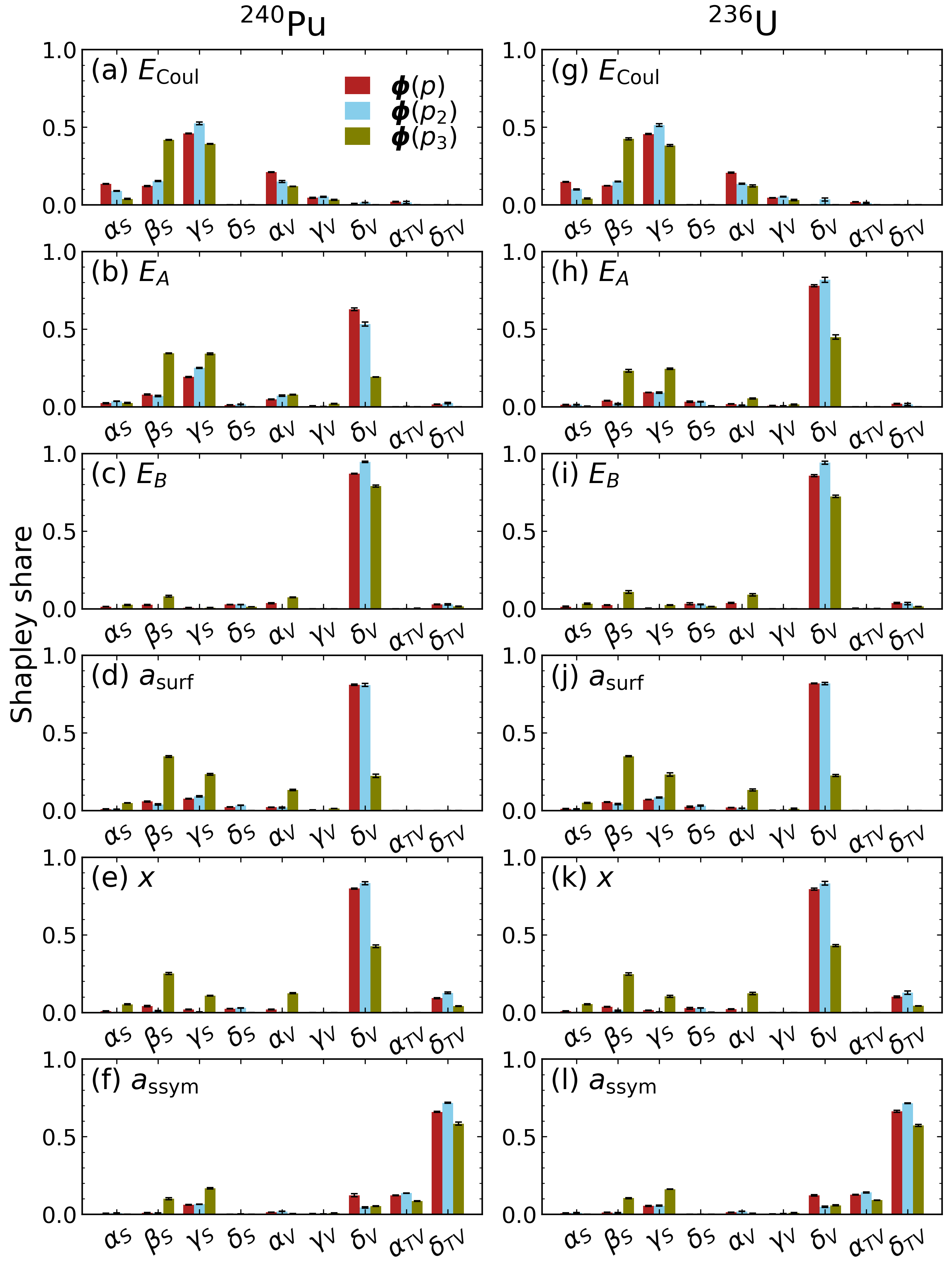}
\caption{Shapley shares of the EDF coupling constants for the different quantities. These are 
computed for $^{240}$Pu [panels (a-f)] and $^{236}$U [panels (g-l)]. 
Each panel compares the Shapley share evaluated with the three distributions: the red bar is with the full posterior distribution $p(\vec{C}|\mathcal{D},\vec{Q}^{\rm ref})$; the blue bar is with the  partially factorized distribution $p_2(\vec{C}|\mathcal{D},\vec{Q}^{\rm ref})$, where $\delta_V$ is decorrelated from the remaining eight coupling constants; and the olive bar is with the fully factorized distribution $p_3(\vec{C}|\mathcal{D},\vec{Q}^{\rm ref})$, constructed as the product of the marginal posterior distribution of each coupling constant. The error bars indicate the standard deviation of the Shapley estimates over $R=64$ bootstrap resamplings of the distribution samples.}
\label{fig:sens}
\end{figure}
Figure~\ref{fig:sens} shows the Shapley shares $\vec{\phi}$ for the selected quantities, computed under three different assumptions for the joint distribution of the coupling constants. The results for \nuc{Pu}{240} and \nuc{U}{236} are shown in Fig.~\ref{fig:sens} (a-f) and Fig.~\ref{fig:sens} (g-l), respectively. The Shapley shares of the coupling constants are normalized such that their sum is unity, 
that is, 
\begin{equation}
    \sum_i\phi_i=1. 
\end{equation}
The error bar on each sensitivity index reflect the statistical uncertainty of the Shapley share due to the finite number of posterior samples, which is estimated by bootstrap resampling of the posterior samples. The red, blue and olive bars are calculated from $p(\vec{C}|\mathcal{D},\vec{Q}^{\rm ref})$, $p_{2}(\vec{C}|\mathcal{D},\vec{Q}^{\rm ref})$ and $p_{3}(\vec{C}|\mathcal{D},\vec{Q}^{\rm ref})$, respectively.
The Shapley shares
are similar for the two nuclei across all quantities and can be summarized as follows:
\begin{itemize}
    \item For the outer barrier heights $E_B$ shown in Figs.~\ref{fig:sens} (c) and (i), the Shapley share of $\delta_V$ closely agrees among $\vec{\phi}(p)$, $\vec{\phi}(p_2)$, and $\vec{\phi}(p_3)$. This indicates that $\delta_V$ remains the dominant coupling constant for the variation of $E_B$ regardless of whether the correlations among the coupling constants are retained or removed. This indicates that $\delta_V$ provides the dominant contribution to the variation of $E_A$, and therefore plays the dominant role in its correlations with the other quantities. 
    
    \item For $E_A$, $a_{\rm surf}$, and $x$ shown in Figs.~\ref{fig:sens} (b, d, e) and (h, j, k), 
    $\vec{\phi}(p)$ of $\delta_V$ agrees closely with $\vec{\phi}(p_2)$, 
    but drops substantially for $\vec{\phi}(p_3)$.  
    The agreement between $\vec{\phi}(p)$ and $\vec{\phi}(p_2)$ shows 
    that the correlation between $\delta_V$ and the remaining eight couplings has little effect on its Shapley share, so that the value obtained from the full posterior already reflects a genuine functional sensitivity. The further reduction from $\vec{\phi}(p_2)$ to $\vec{\phi}(p_3)$ instead reflects the correlations among the remaining eight couplings themselves: once these correlations are removed, their combined contribution is redistributed among several individually correlated parameters, reducing the share attributed to $\delta_V$ alone. Notably,  the combined Shapley share of $\delta_V$, $\beta_S$, $\gamma_S$, and $\alpha_V$ exceeds 90\% under all the three distributions, indicating that the variance of these quantities is governed by these four coupling constants. The correlations between these quantities arise because they vary in a similar manner in response to changes in these coupling constants.
    
    \item  For the Coulomb energy $E_{\rm Coul}$ shown in Figs.~\ref{fig:sens} (a) and (g), the Shapley share of $\delta_V$ is negligible under all the three distributions, in contrast to $a_{\rm surf}$ and the fission quantities. This reflects the fact that $E_{\rm Coul}$ depends on the proton density distribution rather than on the surface stiffness, and is therefore governed by a different combination of the couplings. Although $E_{\rm Coul}$ depends explicitly only on the proton vector density $\rho_V^{(p)}$ as shown in Eq.~(\ref{eq:Coul}), this density is obtained from a self-consistent solution of the Dirac equation, Eq.~(\ref{eq:DiracEq}). As a result, all the coupling constants can in principle influence the proton vector density distribution, and hence $E_{\rm Coul}$. This is in contrast to the case of uniform nuclear matter, where the derivative couplings vanish identically and cannot contribute at all. 
    \item For the surface symmetry energy coefficient $a_{\rm ssym}$ shown in Figs.~\ref{fig:sens} (f) and (l), $\delta_{TV}$ is the dominant coupling under all the three distributions, while the Shapley share of $\delta_V$ remains negligible throughout. This is consistent with the conclusion drawn above that $a_{\rm ssym}$ is primarily determined by the isovector channel of the functional.
\end{itemize}
The same Shapley GSA for the nuclear matter properties are presented in  Fig.~S4 in Supplemental Material.

\begin{figure}[htbp]
    \centering
    \includegraphics[width=0.98\linewidth]{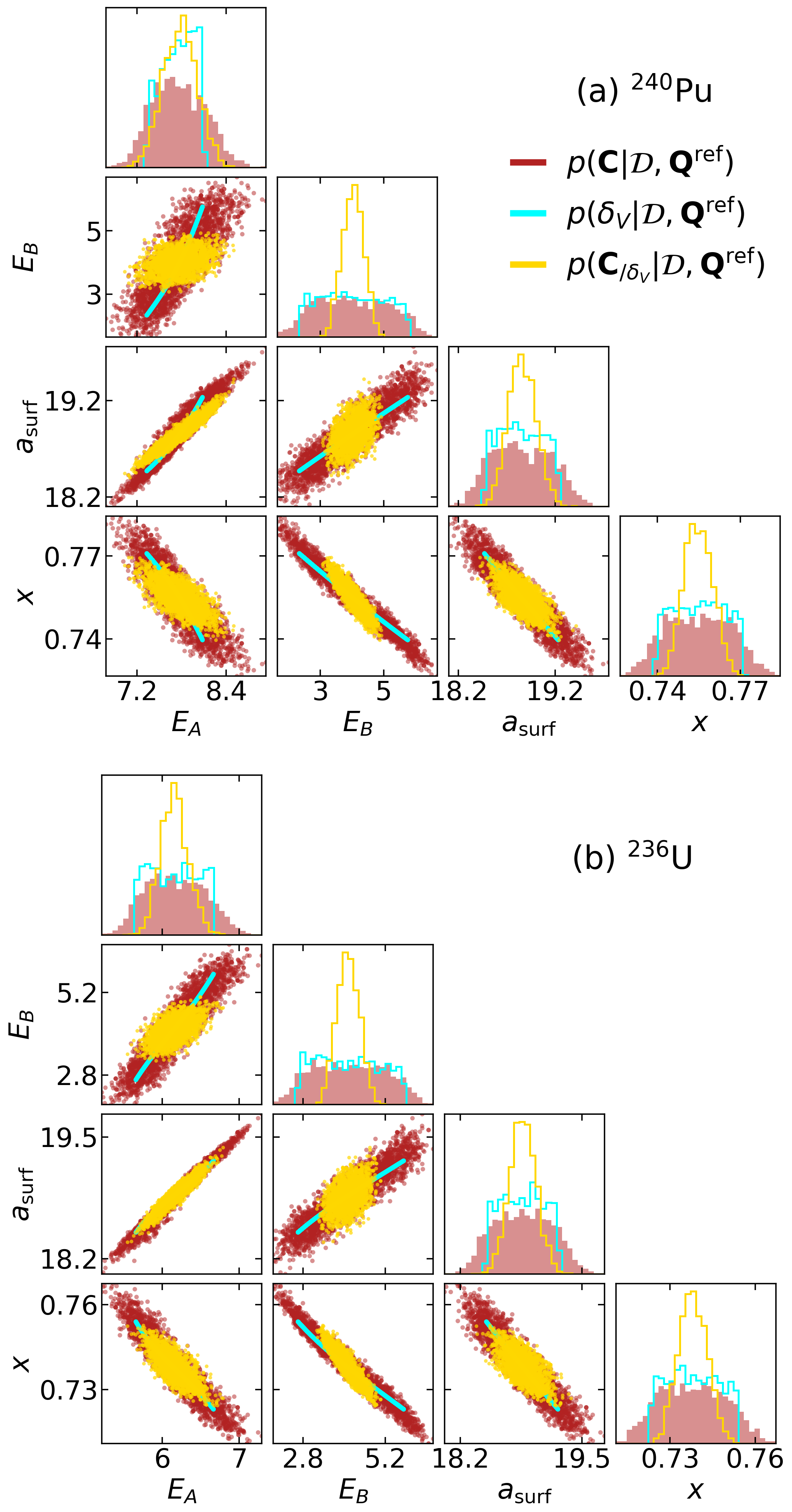}
    \caption{The correlations between the fission barrier heights $E_A$, $E_B$, the surface energy coefficient $a_{\rm surf}$, and the fissility parameter $x$ for $^{240}$Pu (a) and $^{236}$U (b). The points and the histograms are samples drawn from the full posterior $p(\vec{C}|\mathcal{D},\vec{Q}^{\rm ref})$ (red), the marginal posterior $p(\delta_V|\mathcal{D},\vec{Q}^{\rm ref})$ with the remaining eight couplings fixed (cyan), and the marginal posterior $p(\vec{C}_{/\delta_V}|\mathcal{D},\vec{Q}^{\rm ref})$ with $\delta_V$ fixed (yellow). The units of these quantities are the same as in Table~\ref{tab:summary}.}
    \label{fig:corner_dV}
\end{figure}
To further confirm the effect of $\delta_V$, Figs.~\ref{fig:corner_dV} (a) and (b) compare the correlations obtained from the three different distributions for \nuc{Pu}{240} and \nuc{U}{236}, respectively. The red scatters and the histograms are obtained from the full posterior distribution $p(\vec{C}|\mathcal{D},\vec{Q}^{\rm ref})$, while the cyan scatters and the histograms are obtained by varying only $\delta_V$ according to its marginal distribution $p(\delta_V|\mathcal{D},\vec{Q}^{\rm ref})$, with the remaining eight couplings fixed as their PC-PK1 values. On the other hand, the yellow scatters and the histograms are obtained by varying the remaining eight couplings according to their marginal distribution $p(\vec{C}_{/\delta_V}|\mathcal{D},\vec{Q}^{\rm ref})$, with $\delta_V$ fixed at its PC-PK1 value. As can be seen in Fig.~\ref{fig:corner_dV}, varying $\delta_V$ alone reproduces both the trend and the spread of the correlations between $E_B$ with $a_{\rm surf}$ and $x$, while the contribution from varying the remaining eight couplings alone is comparatively small. This confirms that $\delta_V$ is the dominant parameter governing these correlations. In contract, for the inner barrier height $E_A$, both $\delta_V$ and the remaining eight couplings contribute appreciably to the correlation with $a_{\rm surf}$ and $x$, indicating that $E_A$ is jointly governed by $\delta_V$ and the correlated couplings of the functional.

\section{Conclusions}
\label{sec:summary}
In this work, we investigated the correlation between the surface energy coefficients and the fission properties of two representative actinide nuclei, \nuc{Pu}{240} and \nuc{U}{236}, and identified the microscopic origin of this correlation through sensitivity analyses. 
To this end, 
we applied the relativistic Thomas-Fermi approximation within the covariant point-coupling framework and extracted the surface energy coefficient from semi-infinite nuclear matter. Gaussian-process emulators were employed for the statistical analysis, and the posterior distributions of the relevant quantities, together with their statistical uncertainties, were obtained under the  constraint on the ground state properties of \nuc{Pu}{240} and \nuc{U}{236}. The same conclusions for the statistical analysis were obtained when the posterior is constrained by the saturation properties of infinite nuclear matter.

From the posterior distributions, we found that the surface energy coefficient $a_{\rm surf}$ is strongly correlated with both the fission barrier heights and the fissility parameter $x$, whereas the surface symmetry energy coefficient $a_{\rm ssym}$ and the Coulomb energy $E_{\rm Coul}$ exhibit only weak correlations with the fission-related quantities. A Shapley sensitivity analysis further showed that the outer barrier is predominantly sensitive to the isoscalar-vector derivative coupling $\delta_V$, whereas the inner barrier and $a_{\rm surf}$ are jointly governed by $\delta_V$, $\beta_S$, $\gamma_S$, and $\alpha_V$, together with the correlations among these coupling constants. In contrast, $E_{\rm Coul}$ showed little sensitivity to $\delta_V$, while $a_{\rm ssym}$ is instead dominated by the isovector derivative coupling $\delta_{TV}$, which is consistent with their weak correlation with the fission-related quantities. Based on these results, we concluded that the correlation between the outer barrier and the other quantities originates predominantly from $\delta_V$, while the correlation involving the inner barrier arises from $\delta_V$ together with the remaining coupling constants and their correlations. This interpretation was further confirmed by the analysis using the marginalized distributions.

Since fission properties can be constrained by experimental data, the strong correlation between the fission properties and \(a_{\rm surf}\) provides a possible approach to constrain the surface energy coefficient \(a_{\rm surf}\). A more precise determination of \(a_{\rm surf}\) could improve 
the treatment of surface effects in systems relevant to nuclear astrophysics. For example, \(a_{\rm surf}\)  plays an important role in determining the shape transitions in the inner crust of neutron stars, where complex nuclear configurations known as nuclear pasta may emerge~\cite{Bao2014,refId0}. Moreover, because we found that the fission properties, especially $E_B$, are sensitive to \(\delta_V\), experimental information on fission observables, such as spontaneous-fission half-lives, could also be incorporated to constrain \(\delta_V\).

In this paper, we considered only axially-symmetric deformations. It would be an interesting future work 
to extend the present framework by including the triaxial deformation degree of freedom into the calculation of fission barrier and by exploring a broader set of spontaneously fissioning nuclei.

\section*{Acknowledgments} 
 We thank  H. Togashi and S. Yoshikawa for fruitful discussions. This work was supported by JSPS KAKENHI Grant Numbers JP23K03414, 26H01416 and JZ08699. The numerical calculations were performed with the computer facility at the Yukawa Institute for Theoretical Physics, Kyoto University.

\appendix
\section{Derivation of the Surface Tension in Semi-Infinite Nuclear Matter}
\label{app:sinm}

In semi-infinite nuclear matter (SINM), 
the density varies along a
single direction $z$, interpolating between saturated bulk matter as
$z\to-\infty$ and the vacuum as $z\to+\infty$. 
Throughout this Appendix, the asymmetry $I$ is held fixed, and we omit its explicit dependence in the energy functional for nuclear matter, $\mathcal{E}(\rho,I)$ in Eq. (\ref{eq:energy}), the excess energy density, $w(\rho,I)$ in Eq. (\ref{eq:w}), and the stiffness function $C(\rho,I)$ in Eq.~(\ref{eq:C}),  writing them simply as $\mathcal{E}(\rho)$, $w(\rho)$, $C(\rho)$, etc. All the derivatives with respect to $\rho$ below are therefore understood to be taken at fixed $I$. The energy of the SINM slab can then be expressed as
\begin{align}
\label{eq:ESINM}
  E_{\rm SINM}
  &=\int\!\Big[\mathcal E(\rho)+C(\rho)\,(\nabla\rho)^2\Big]\,d^3r\\
  & =S\!\int\!\Big[\mathcal E(\rho)+C(\rho)\,\rho'^{2}\Big]\,dz ,
\end{align}
where $S$ is the interface area, $\rho'\equiv d\rho/dz$, $\mathcal E(\rho)$ is
the local (Thomas--Fermi) energy density of uniform matter, and
$C(\rho)\,\rho'^{2}$ is the derivative energy arising from the
finite-range couplings of the functional. The explicit form of $C(\rho)$ is
derived at the end of this appendix.

The surface tension $\sigma$ is the excess energy per unit area of the SINM slab relative
to a reference state of saturated bulk matter occupying the same volume. Because
the slab is depleted of particles in the surface region, the two systems, that is, the SINM and the bulk matter, do not contain the same number of nucleons. A meaningful comparison must restore this
difference by $\mu\,\Delta N$, where $\mu$ is introduced as the Lagrange multiplier associated with particle-number conservation. Hence
\begin{equation}
\label{eq:sigma_def}
  \sigma=\frac{E_{\rm SINM}-E_{\rm bulk}}{S}
        +\mu\,\frac{N_{\rm bulk}-N_{\rm SINM}}{S},
\end{equation}
where $N_{\rm bulk}$ and $N_{\rm SINM}$ are the particle numbers of the
reference bulk matter and of the SINM slab, respectively, and
$E_{\rm bulk}=S\!\int\mathcal E(\rho_0)\,dz$. 
At the equilibrium, the chemical potential $\mu$ is defined by Eq.~(\ref{eq:chemical_pot}).
With $N_{\rm SINM}/S=\int\rho\,dz$ and $N_{\rm bulk}/S=\int\rho_0\,dz$ and
collecting the integrands, Eq.~\eqref{eq:sigma_def} becomes
\begin{equation}
\label{eq:sigma_raw}
  \sigma=\int_{-\infty}^{\infty}\!
  \Big[\,C(\rho)\,\rho'^{2}
        +\mathcal E(\rho)-\mathcal E(\rho_0)
        -\mu\big(\rho-\rho_0\big)\,\Big]\,dz .
\end{equation}
The saturation condition of bulk matter leads to zero pressure at $\rho_0$,
\begin{equation}
\label{eq:sat}
  P(\rho_0)=\mu\rho_0-\mathcal E(\rho_0)=0 ,
\end{equation}
which yields
\begin{align}
\label{eq:mu}
\mu=\frac{\mathcal E(\rho_0)}{\rho_0}
\end{align}
and 
\begin{align}
\label{eq:sigma}
  \sigma=\int_{-\infty}^{\infty}\!
\Big[\,C(\rho)\,\rho'^{2}+w(\rho)\,\Big]\,dz ,
\end{align}
where 
\begin{align}
    w(\rho)\equiv\mathcal E(\rho)-\mu\rho ,
\end{align}
is the excess grand-potential density measured relative to the bulk equilibrium phase.

Requiring $\delta\sigma=0$ with a variation of the density, $\rho\to\rho+\delta\rho$, yields the Euler--Lagrange
equation
\begin{equation}
\label{eq:EL}
  C'(\rho)\,\rho'^{2}+2C(\rho)\,\rho''=\frac{d\mathcal E}{d\rho}-\mu ,
\end{equation}
which, upon multiplication by $\rho'$, is the total derivative,
\begin{equation}
\label{eq:total}
  \frac{d}{dz}\!\big[C(\rho)\,\rho'^{2}\big]
  =\frac{d}{dz}\!\big[\mathcal E(\rho)-\mu\rho\big]
  =\frac{dw}{dz} .
\end{equation}
Integrating once gives $C(\rho)\,\rho'^{2}=w(\rho)+\text{const}$. The constant is
fixed to be zero with the boundary condition of $w\to0$ for $\rho'\to0$. 
This yields
the equipartition between gradient and potential energy characteristic of
square-gradient theories,
\begin{equation}
\label{eq:equip}
  C(\rho)\,\rho'^{2}=w(\rho) .
\end{equation}
Inserting Eq.~\eqref{eq:equip} into Eq.~\eqref{eq:sigma} replaces the gradient
term by $w$, so that
\begin{equation}
\label{eq:sigma2}
  \sigma=2\int_{-\infty}^{\infty}\!w(\rho)\,dz .
\end{equation}
Equation~\eqref{eq:equip} also gives $\rho'=\sqrt{w/C}$, hence
$dz=\sqrt{C/w}\,d\rho$. Changing the integration variable from $z$ to $\rho$
eliminates the profile entirely, and the integration limits become the vacuum
($\rho=0$) and the bulk ($\rho=\rho_0$):
\begin{equation}
\label{eq:sigma3}
\sigma(I)=2\int_{0}^{\rho_0}\!\sqrt{C(\rho)\,w(\rho)}\;d\rho . \;
\end{equation}
The coefficient $C(\rho)$ follows from the gradient (derivative-coupling) part of
the functional. From the point-coupling Lagrangian given by Eq.~\eqref{e2}, the
gradient energy reads
\begin{equation}
\label{eq:Egrad_raw}
  E_{\rm grad}=\frac12\int\!\Big[
  \delta_S\,\rho_S\,\Delta\rho_S
  +\delta_V\,\rho_V\,\Delta\rho_V
  +\delta_{TV}\,\vec\rho_{TV}\!\cdot\!\Delta\vec\rho_{TV}\Big]\,d^3r .
\end{equation}
In SINM we take the baryon density $\rho_V$ as the independent variable, because  $\rho_S$ is obtained from $\rho_V$ by the  relation introduced in Sec.~\ref{sc:SINM} and the isovector density follows $\rho_{TV}=I\rho_V$. Using these relations and
integrating by parts, we obtain
\begin{equation}
\label{eq:Egrad}
  E_{\rm grad}=-\frac{S}{2}\int\!\Big[
  \delta_V
  +\delta_S\Big(\frac{d\rho_S}{d\rho_V}\Big)^{\!2}
  +\delta_{TV}\,I^{2}\Big](\rho_V')^{2}\,dz .
\end{equation}
Comparison with the gradient term $C(\rho)\,\rho'^{2}$ of Eq.~\eqref{eq:ESINM}
identifies
\begin{equation}
\label{eq:C_appendix}
 \;
  C(\rho)=-\frac12\Big[
  \delta_V
  +\delta_S\Big(\frac{d\rho_S}{d\rho_V}\Big)^{\!2}
  +\delta_{TV}\,I^{2}\Big] .
\end{equation}

\section{Benchmark of the Gaussian-process emulator}
\label{app:benchmark}
\begin{figure}[htbp]
\includegraphics[width=\columnwidth]{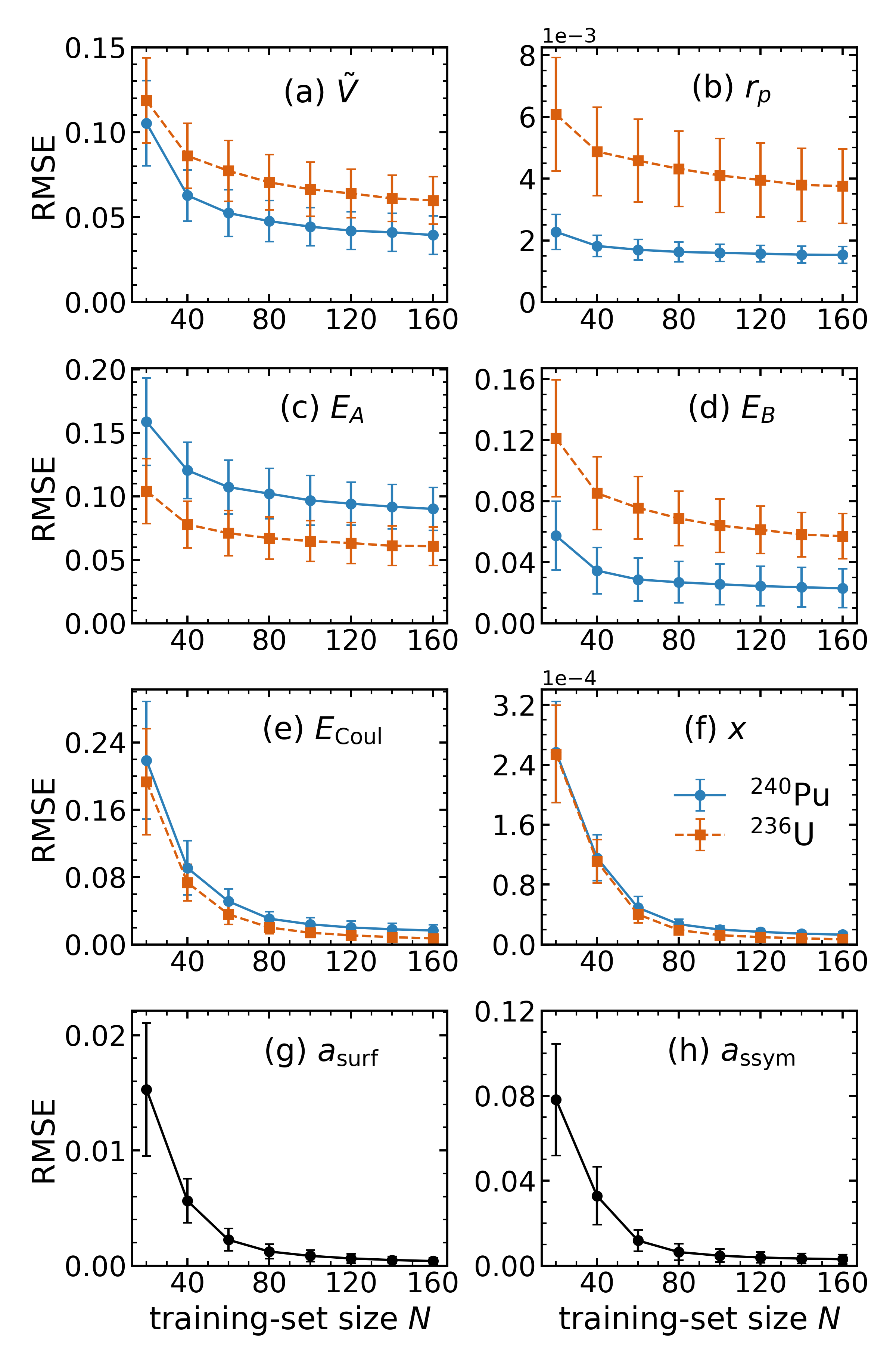}
\caption{The root-mean-square-error (RMSE) of the predictions for different quantities calculated from the GP emulators as a function of the training-set size $N$. The RMSE value for each $N$ is evaluated by Monte Carlo cross-validation with $R=64$ independent random partitions. For each $N$, the emulator is trained on $N$ randomly selected samples and tested on random 20 testing set selected from the rest samples. The plotted value is the mean RMSE over all 64 repetitions and the errorbars indicate the standard deviation across all repetitions. In the panel (a), $\tilde{V}$ is defined as $\tilde{V} = \sqrt{\frac{1}{M}\sum_{j=1}^{M}\left[V^{\rm pred}(\beta_{2}^{(j)}) - V^{\rm test}(\beta_{2}^{(j)})\right]^2}$ where $ V^{\rm pred}(\bt)$ is the predicted energy potential curve and $V^{\rm test}(\bt)$ is the energy potential curve of the corresponding testing set. The unit of $r_p$ is fm, the fissility parameter $x$ is dimensionless, and all the other quantities are given in MeV.  Results for \nuc{Pu}{240} and \nuc{U}{236} are shown in the blue circles and the orange squares, respectively. The black solid line shows the RMSE for the surface energy coefficient $a_{\rm surf}$ and the surface symmetry energy coefficient $a_{\rm ssym}$ from SINM.}
\label{fig:learning}
\end{figure}

To assess the predictive accuracy of the Gaussian-process (GP) emulators, we perform a benchmark using the 200 EDF parameter sets introduced in Sec.~\ref{sec:emulator}.  Fig.~\ref{fig:learning} shows the benchmark results of the GP emulators for different quantities for \nuc{Pu}{240} (the solid blue line) and \nuc{U}{236} (the dashed orange line) as well as SINM (the solid black line). The value and uncertainty for each point is obtained by Monte Carlo cross-validation. For each training-set size, a subset of $N$ parameter sets is drawn at random from the full samples as the training set. From the remaining samples, 20 samples are selected at random to form the testing set.  This random partition is repeated independently 64 times. For each training-set size, each point represents the mean root-mean-square error (RMSE) over the 64 repetitions, while the error bar corresponds to the standard deviation of the RMSE values. $\tilde{V}$ in Fig.~\ref{fig:learning} (a) is defined as $\tilde{V}\equiv\sqrt{\frac{1}{M}\sum_{j=1}^{M}\left[V^{\rm pred}(\beta_{2}^{(j)}) - V^{\rm test}(\beta_{2}^{(j)})\right]^2}$, where $ V^{\rm pred}(\bt)$ is the predicted PEC by PCA-GP and $V^{\rm test}(\bt)$ is the PEC of the corresponding testing set. Here, $\{\beta_2^{(j)}\}_{j=1}^{M}$ denotes the $M$ mesh-points of the common densified grid for the quadrupole deformation. For all the quantities, the learning curves exhibit a steep decrease initially,  followed by a gradual plateau after $N=120$. In Fig.~\ref{fig:learning} (a) for the RMSE of the full PEC, even though the curve has not fully been converged, the reduction of the mean RMSE from $N=140$ to $N=160$ is less than $0.004$~MeV, well below the energy scale of the PEC itself. 

\newpage
\bibliographystyle{apsrev4-1} 
 \bibliography{ref}

\end{document}


\newcommand{\BEhead}[2]{%
\begin{tabular}[c]{@{}c@{}}
$B(E2;$\\
$#1\!\to\!#2)$
\end{tabular}}

\title{Supplemental Material}
\author{}
\maketitle

\setcounter{section}{0}
\setcounter{equation}{0}
\setcounter{figure}{0}
\setcounter{table}{0}
\renewcommand{\theequation}{S\arabic{equation}}
\renewcommand{\thefigure}{S\arabic{figure}}
\renewcommand{\thetable}{S\arabic{table}}

\renewcommand{\vec}[1]{\mbox{\boldmath $#1$}}

\section{Additional Results for the Posterior Constrained by $\vec{Q}^{\rm ref}_1$}
\label{sec:posterior_O1}
\begin{figure*}[t]
    \centering 
    \includegraphics[width=2\columnwidth]{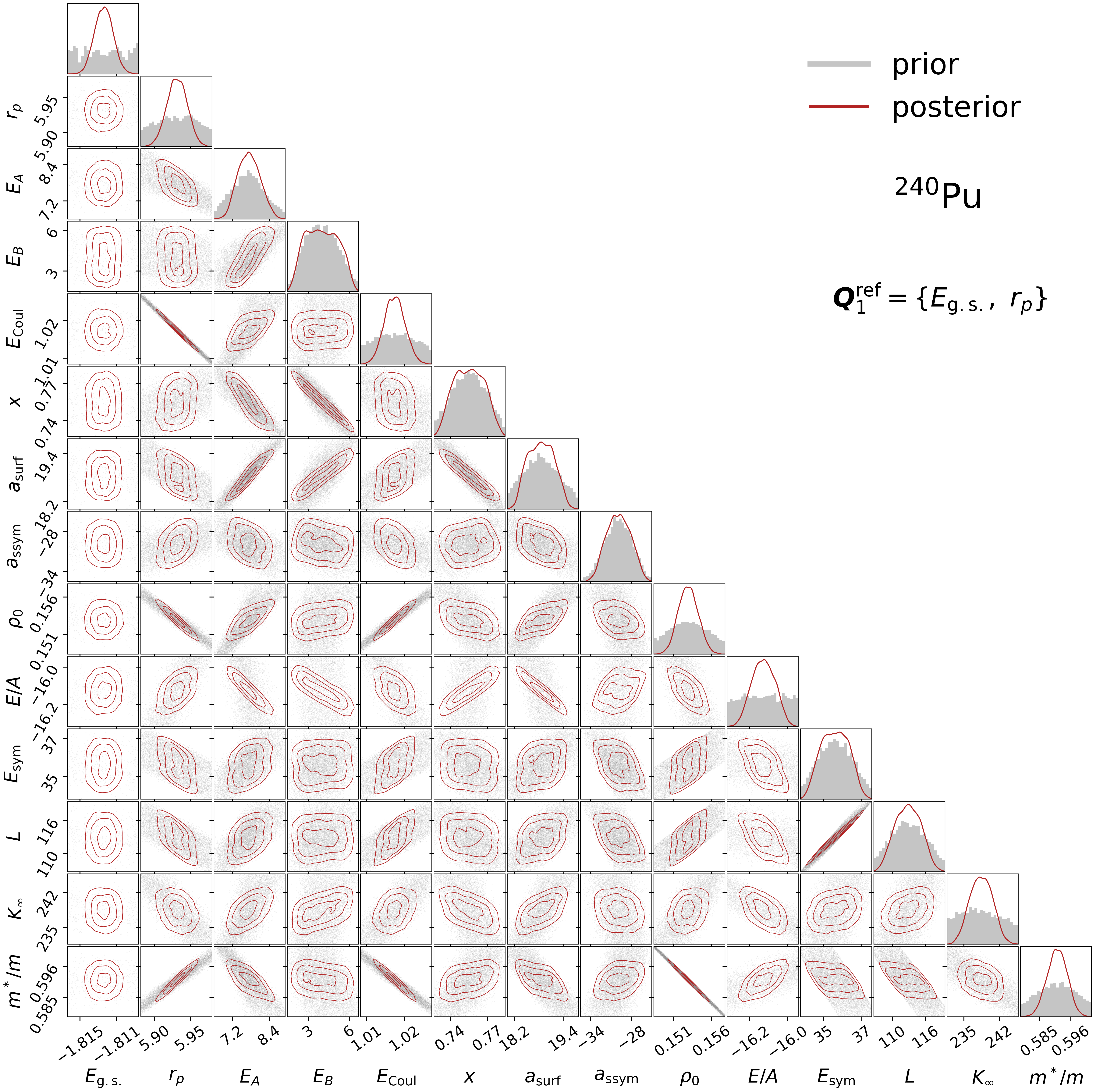}
    \caption{Same as Fig.~3 in the main text, but including all the quantities of $^{240}$Pu and the nuclear matter properties. The units of $E_{\rm g.s.}$ and $E_{\rm Coul}$ are GeV, while the units of all the other quantities are the same as those listed in Table~\ref{tab:summary1}.}
    \label{fig-1}
\end{figure*}

\begin{figure*}[t]
    \centering 
    \includegraphics[width=2\columnwidth]{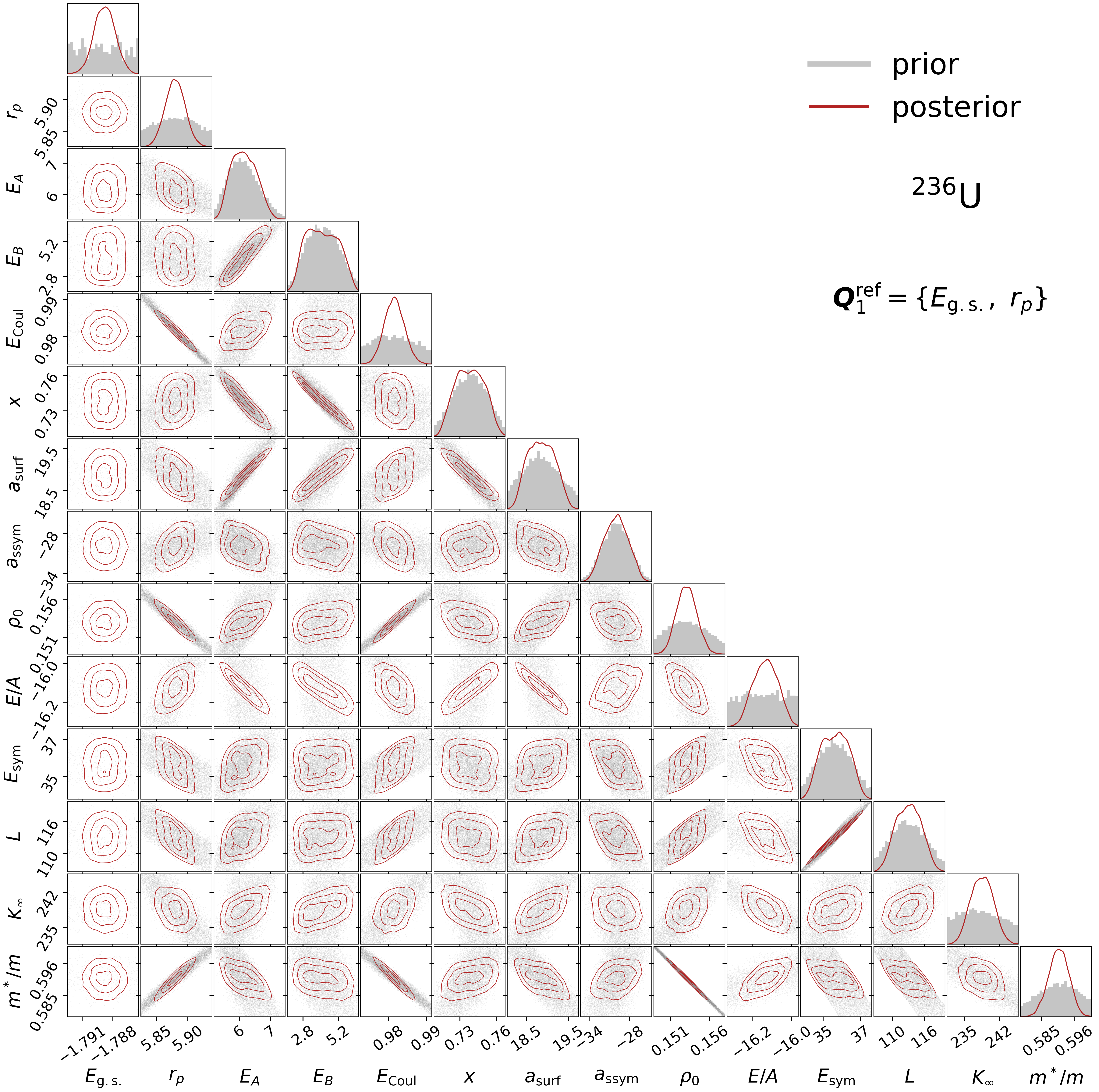}
    \caption{Same as Fig.~\ref{fig-1}, but for $^{236}$U.}
    \label{fig-2}
\end{figure*}

\begin{figure}[!htbp]
    \centering 
    \includegraphics[width=.95\columnwidth]{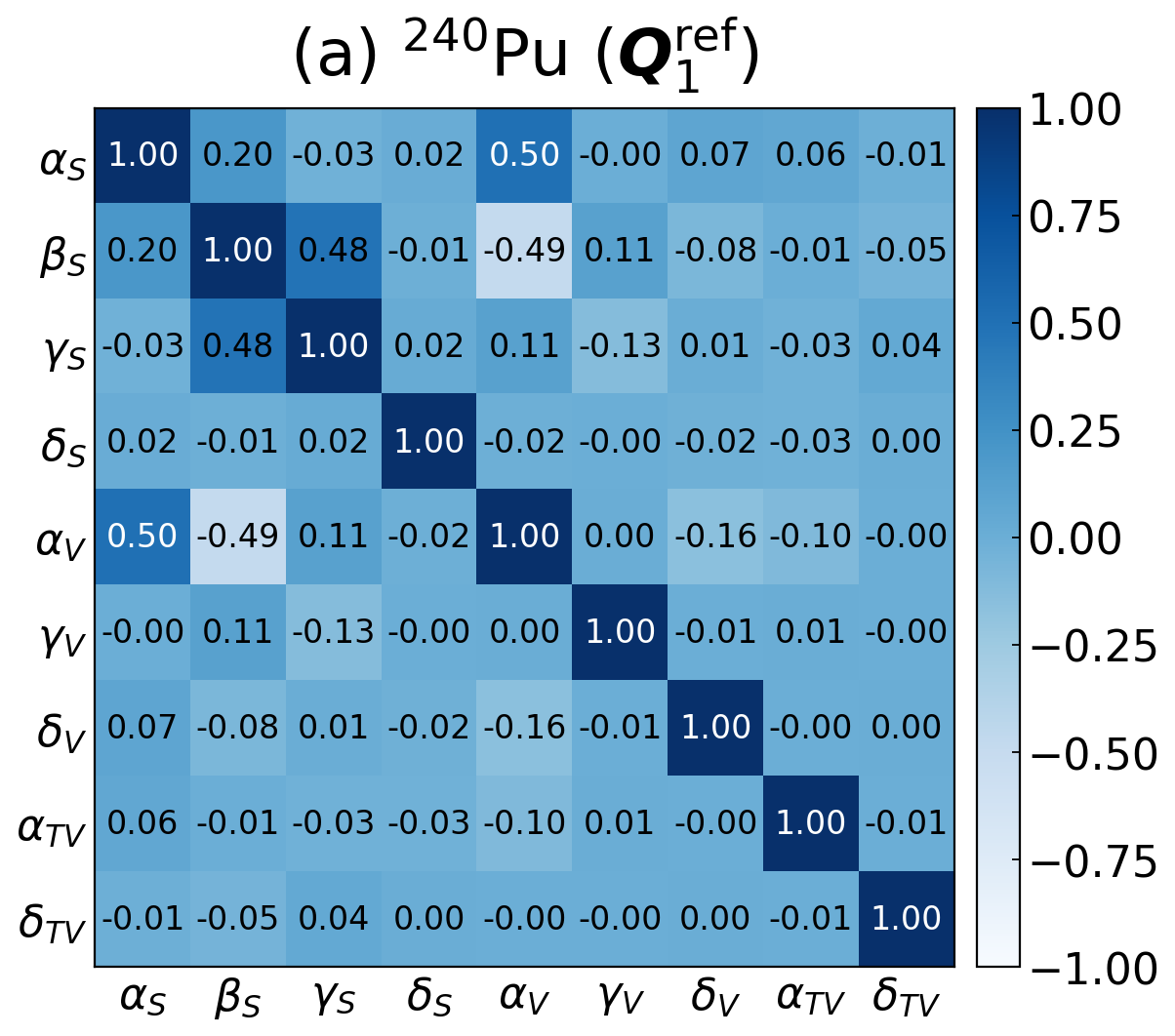}
    
    \vspace*{0.7cm}
    
    \includegraphics[width=.95\columnwidth]{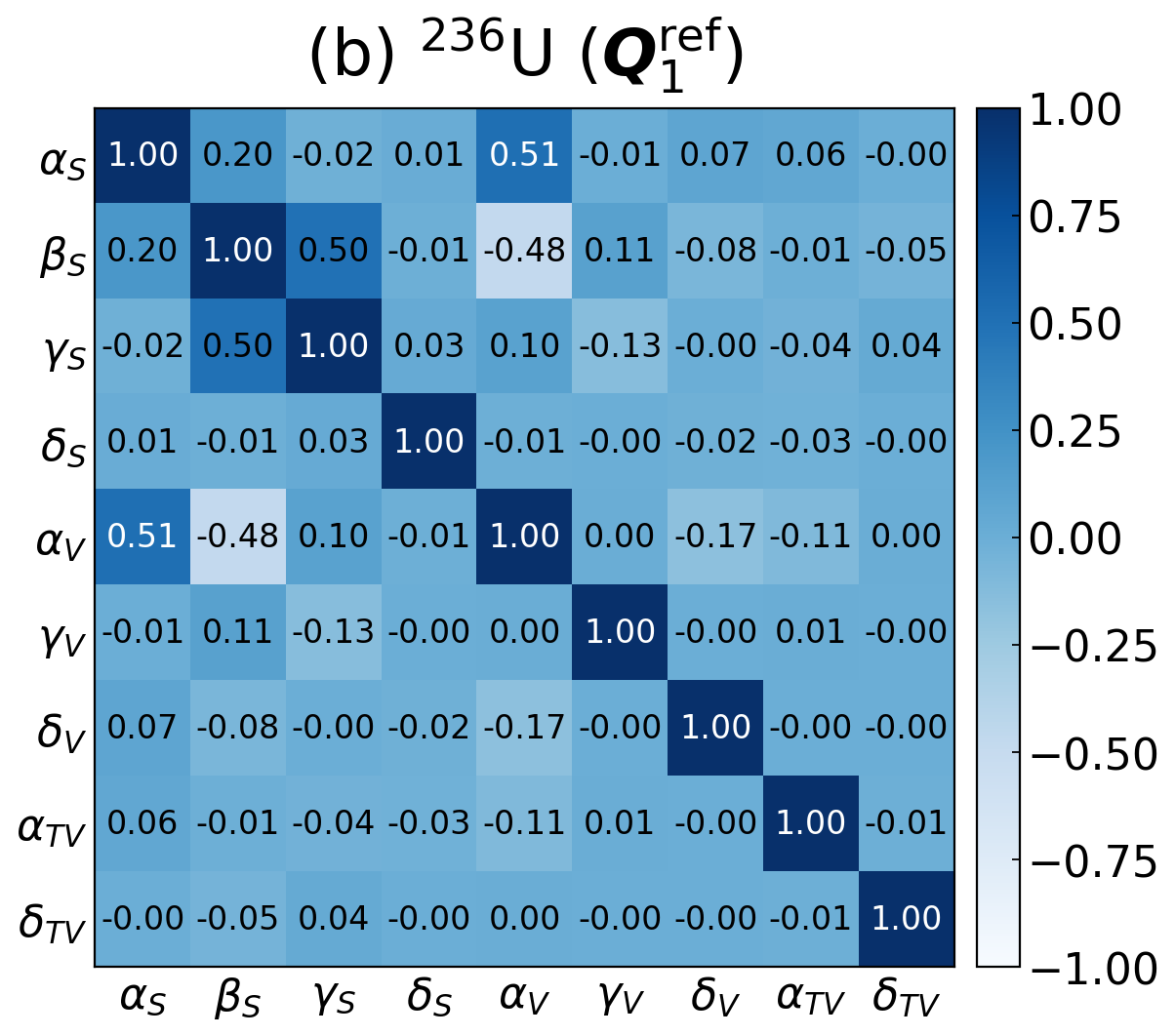}
    \caption{Pearson correlation coefficients among the nine coupling constants of the PC-PK1 functional in the posterior distribution constrained by the ground-state properties. Panels (a) and (b) show the results for $^{240}$Pu and $^{236}$U, respectively.}
    \label{fig:para_corre}
\end{figure}

\begin{figure}[!htbp]
    \centering
    \includegraphics[width=1.03\columnwidth]{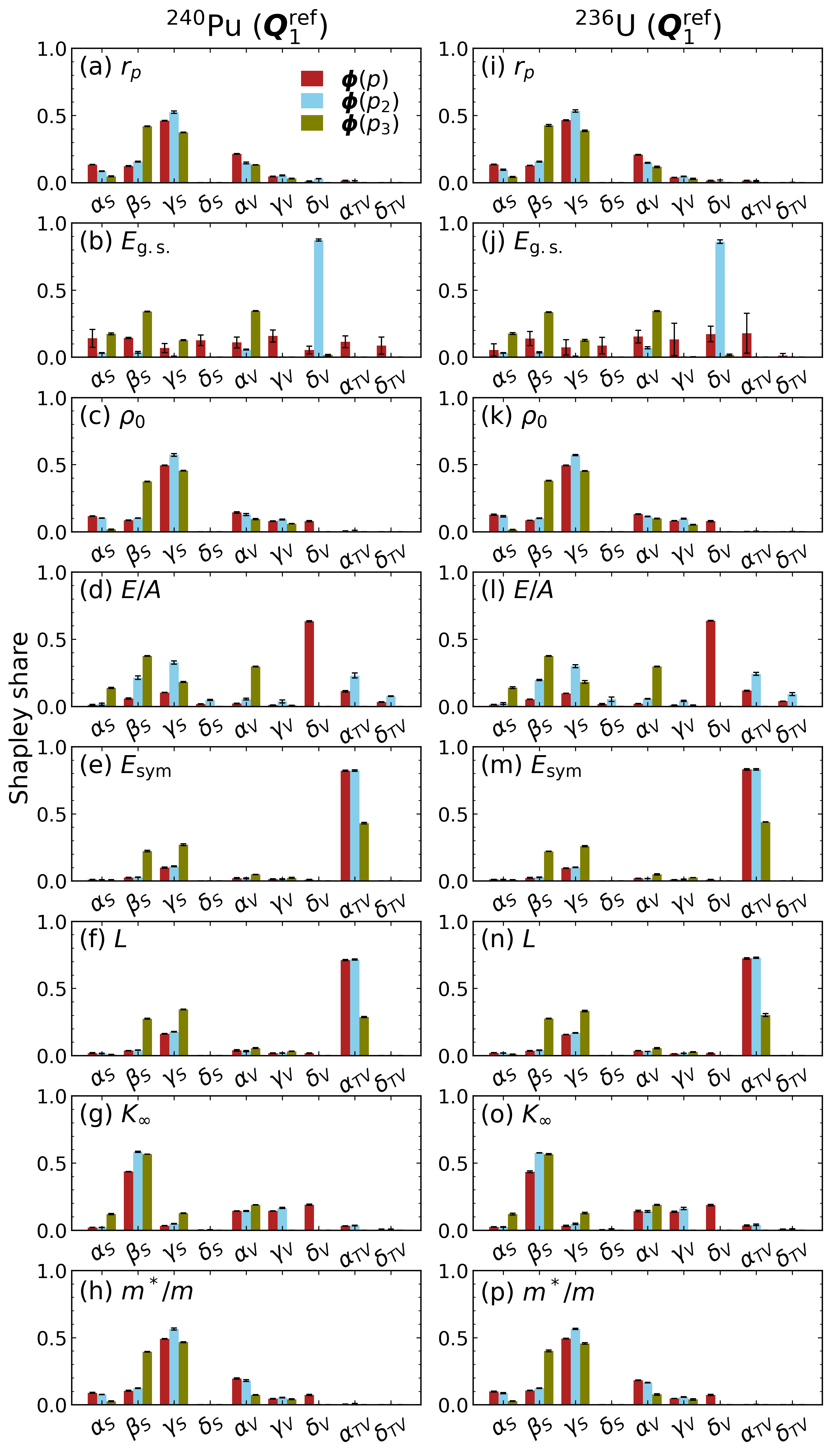}
    \caption{Same as Fig.~4 in the main text, but for different ground-state properties and nuclear matter properties.}
    \label{fig:gs_sens}
\end{figure}

In this section, we present the additional results for the posterior constrained by $\vec{Q}^{\rm ref}_1=\{E_{\rm g.s.}, r_p\}$. Table~\ref{tab:summary1} presents the means and standard deviations of the prior and posterior distributions for all the observables discussed in the main text. The correlations among these observables are shown in Figs.~\ref{fig-1} and \ref{fig-2} for $^{240}\mathrm{Pu}$ and $^{236}\mathrm{U}$, respectively. We also present the correlations among the 
coupling constants of the PC-PK1 functional
from the parameter posterior distribution in Fig.~\ref{fig:para_corre}, as well as their Shapley sensitivity shares for each observable in Fig.~\ref{fig:gs_sens}.

\begin{table}[!htbp]
\centering
\renewcommand{\arraystretch}{1.4}
\caption{Same as Table.~2 in the main text, but including all quantities of $^{240}$Pu, $^{236}$U and the nuclear matter properties.}
\label{tab:summary1}
\begin{tabular}{lll}
\hline\hline
Observable & Prior & Posterior \\
\hline
\multicolumn{3}{c}{$^{240}$Pu} \\
\hline
$E_{\rm g.s.}$ (MeV)   & $-1811.462\,(55.214)$ & $-1812.368\,(1.010)$ \\
$r_p$ (fm)             & $5.932\,(0.045)$      & $5.932\,(0.014)$ \\
$E_A$ (MeV)            & $7.700\,(0.592)$      & $7.755\,(0.355)$ \\
$E_B$ (MeV)            & $4.014\,(1.203)$      & $4.029\,(1.121)$ \\
$E_{\rm C}$ (MeV)      & $1017.571\,(7.981)$   & $1017.454\,(2.638)$ \\
$x$                    & $0.755\,(0.015)$      & $0.755\,(0.012)$ \\
$a_{\rm surf}$ (MeV)   & $18.860\,(0.510)$     & $18.852\,(0.303)$ \\
$a_{\rm ssym}$ (MeV)   & $-29.949\,(2.026)$    & $-29.888\,(1.735)$ \\
$\rho_0$ (fm$^{-3}$)   & $0.153\,(0.003)$      & $0.153\,(0.001)$ \\
$E/A$ (MeV)            & $-16.124\,(0.325)$    & $-16.128\,(0.060)$ \\
$E_{\rm sym}$ (MeV)    & $35.623\,(0.987)$     & $35.601\,(0.694)$ \\
$L$ (MeV)              & $112.956\,(3.626)$    & $112.866\,(2.277)$ \\
$K_\infty$ (MeV)              & $238.773\,(6.902)$    & $238.462\,(2.166)$ \\
$m^*/m$                & $0.591\,(0.008)$      & $0.591\,(0.003)$ \\
\hline
\multicolumn{3}{c}{$^{236}$U} \\
\hline
$E_{\rm g.s.}$ (MeV)   & $-1788.033\,(54.186)$ & $-1788.797\,(1.011)$ \\
$r_p$ (fm)             & $5.883\,(0.044)$      & $5.880\,(0.015)$ \\
$E_A$ (MeV)            & $6.139\,(0.489)$      & $6.163\,(0.380)$ \\
$E_B$ (MeV)            & $4.167\,(1.125)$      & $4.137\,(1.036)$ \\
$E_{\rm C}$ (MeV)      & $981.383\,(7.783)$    & $981.577\,(2.545)$ \\
$x$                    & $0.739\,(0.015)$      & $0.738\,(0.012)$ \\
$a_{\rm surf}$ (MeV)   & $18.860\,(0.510)$     & $18.866\,(0.304)$ \\
$a_{\rm ssym}$ (MeV)   & $-29.949\,(2.026)$    & $-29.963\,(1.733)$ \\
$\rho_0$ (fm$^{-3}$)   & $0.153\,(0.003)$      & $0.153\,(0.001)$ \\
$E/A$ (MeV)            & $-16.124\,(0.325)$    & $-16.131\,(0.060)$ \\
$E_{\rm sym}$ (MeV)    & $35.623\,(0.987)$     & $35.653\,(0.696)$ \\
$L$ (MeV)              & $112.956\,(3.626)$    & $113.060\,(2.277)$ \\
$K_\infty$ (MeV)              & $238.774\,(6.902)$    & $238.583\,(2.186)$ \\
$m^*/m$                & $0.591\,(0.008)$      & $0.591\,(0.003)$ \\
\hline\hline
\end{tabular}
\end{table}

\FloatBarrier

\section{Additional Results for the Posterior Constrained by $\vec{Q}^{\rm ref}_2$}

In this section, we present the same results as those presented in Sec.~\ref{sec:posterior_O1}, but for the posterior constrained by  $\vec{Q}^{\rm ref}_2=\{\rho_0, E/A, E_{\rm sym},L, K_{\infty},m^*/m\}$. The reference values for PC-PK1 are $\rho_0=0.153$ fm$^{-3}$, $E/A=-16.119$ MeV, $E_{\rm sym}=35.602$ MeV, $L=112.869$ MeV, $K_\infty=238.181$ MeV, and $m^*/m=0.591$, with the corresponding tolerances of $\sigma_{\rm tol}^{(\rm \rho_0)}=0.004$ fm$^{-3}$, $\sigma_{\rm tol}^{(\rm E/A)}=0.400$ MeV, $\sigma_{\rm tol}^{(E_{\rm sym})}=1.400$ MeV, $\sigma_{\rm tol}^{(L)}=3.000$ MeV, $\sigma_{\rm tol}^{(K_\infty)}=7.000$  MeV, and $\sigma_{\rm tol}^{(m^*/m)}=0.030$, respectively.

\begin{table}[!htbp]
\centering
\renewcommand{\arraystretch}{1.4}
\caption{Same as Table.~\ref{tab:summary1}, but for the posterior distribution constrained by $\vec{Q}^{\rm ref}_2$.}
\label{tab:summary2}
\begin{tabular}{lll}
\hline\hline
Observable & Prior & Posterior \\
\hline
\multicolumn{3}{c}{$^{240}$Pu} \\
\hline
$E_{\rm g.s.}$ (MeV)   & $-1811.462\,(55.214)$ & $-1809.820\,(33.054)$ \\
$r_p$ (fm)             & $5.932\,(0.045)$      & $5.933\,(0.023)$ \\
$E_A$ (MeV)            & $7.700\,(0.592)$      & $7.724\,(0.377)$ \\
$E_B$ (MeV)            & $4.014\,(1.203)$      & $4.014\,(1.104)$ \\
$E_{\rm C}$ (MeV)      & $1017.571\,(7.981)$   & $1017.267\,(4.169)$ \\
$x$                    & $0.755\,(0.015)$      & $0.755\,(0.012)$ \\
$a_{\rm surf}$ (MeV)   & $18.860\,(0.510)$     & $18.844\,(0.322)$ \\
$a_{\rm ssym}$ (MeV)   & $-29.949\,(2.026)$    & $-29.880\,(1.682)$ \\
$\rho_0$ (fm$^{-3}$)   & $0.153\,(0.003)$      & $0.153\,(0.002)$ \\
$E/A$ (MeV)            & $-16.124\,(0.325)$    & $-16.115\,(0.174)$ \\
$E_{\rm sym}$ (MeV)    & $35.623\,(0.987)$     & $35.591\,(0.557)$ \\
$L$ (MeV)              & $112.956\,(3.626)$    & $112.826\,(1.881)$ \\
$K_\infty$ (MeV)              & $238.773\,(6.902)$    & $238.256\,(3.800)$ \\
$m^*/m$                & $0.591\,(0.008)$      & $0.591\,(0.004)$ \\
\hline
\multicolumn{3}{c}{$^{236}$U} \\
\hline
$E_{\rm g.s.}$ (MeV)   & $-1788.033\,(54.186)$ & $-1786.460\,(32.441)$ \\
$r_p$ (fm)             & $5.883\,(0.044)$      & $5.883\,(0.024)$ \\
$E_A$ (MeV)            & $6.139\,(0.489)$      & $6.133\,(0.369)$ \\
$E_B$ (MeV)            & $4.167\,(1.125)$      & $4.130\,(1.011)$ \\
$E_{\rm C}$ (MeV)      & $981.383\,(7.783)$    & $981.097\,(4.082)$ \\
$x$                    & $0.739\,(0.015)$      & $0.739\,(0.012)$ \\
$a_{\rm surf}$ (MeV)   & $18.860\,(0.510)$     & $18.843\,(0.321)$ \\
$a_{\rm ssym}$ (MeV)   & $-29.949\,(2.026)$    & $-29.886\,(1.683)$ \\
$\rho_0$ (fm$^{-3}$)   & $0.153\,(0.003)$      & $0.153\,(0.002)$ \\
$E/A$ (MeV)            & $-16.124\,(0.325)$    & $-16.115\,(0.174)$ \\
$E_{\rm sym}$ (MeV)    & $35.623\,(0.987)$     & $35.592\,(0.557)$ \\
$L$ (MeV)              & $112.956\,(3.626)$    & $112.830\,(1.879)$ \\
$K_\infty$ (MeV)              & $238.774\,(6.902)$    & $238.255\,(3.799)$ \\
$m^*/m$                & $0.591\,(0.008)$      & $0.591\,(0.004)$ \\
\hline\hline
\end{tabular}
\end{table}

\begin{figure*}[t]
    \centering 
    \includegraphics[width=2\columnwidth]{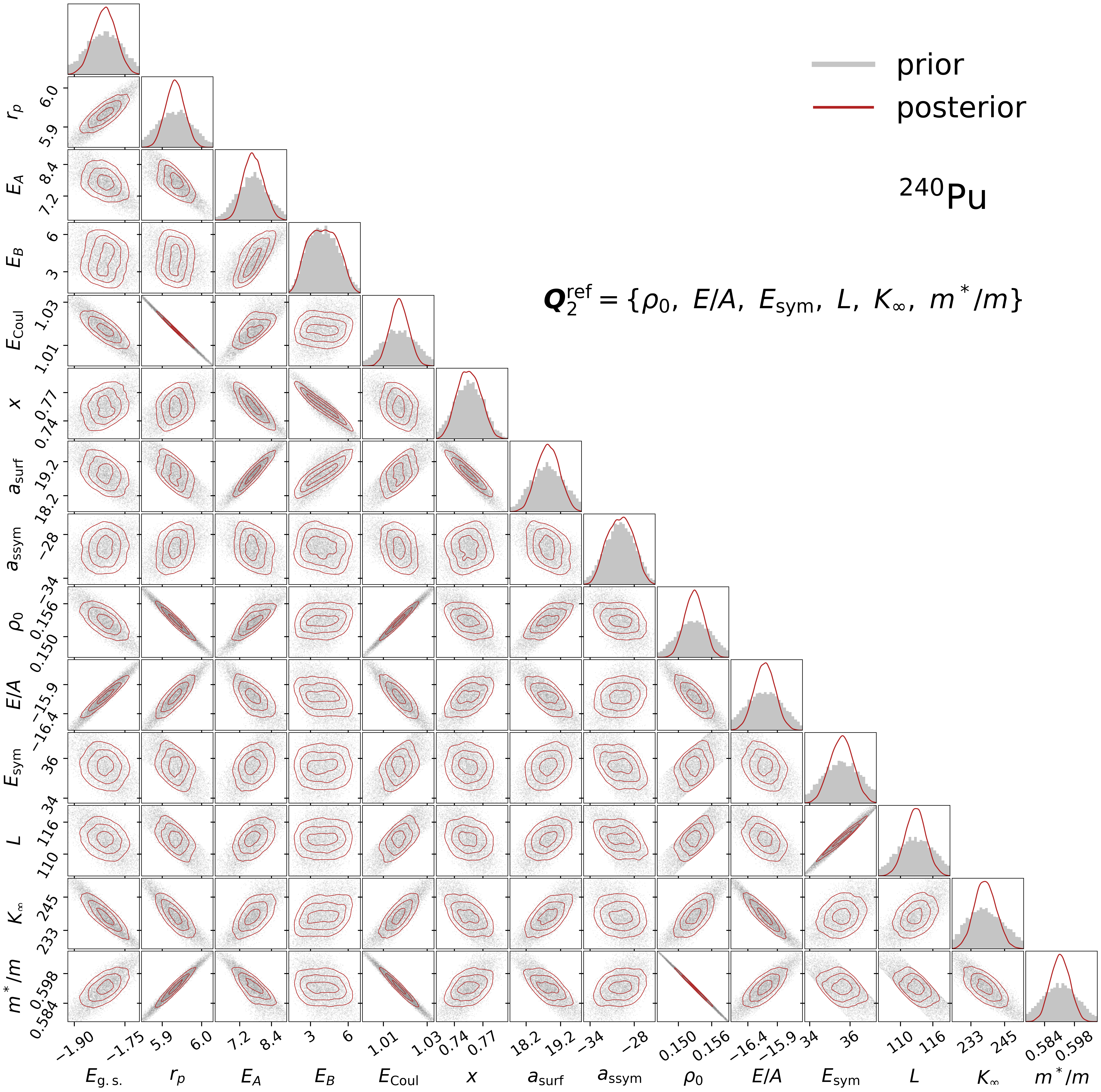}
    \caption{Same as Fig.~\ref{fig-1}, but for the posterior distribution constrained by $\vec{Q}^{\rm ref}_2$.}
    \label{fig-3}
\end{figure*}

\begin{figure*}[t]
    \centering 
    \includegraphics[width=2\columnwidth]{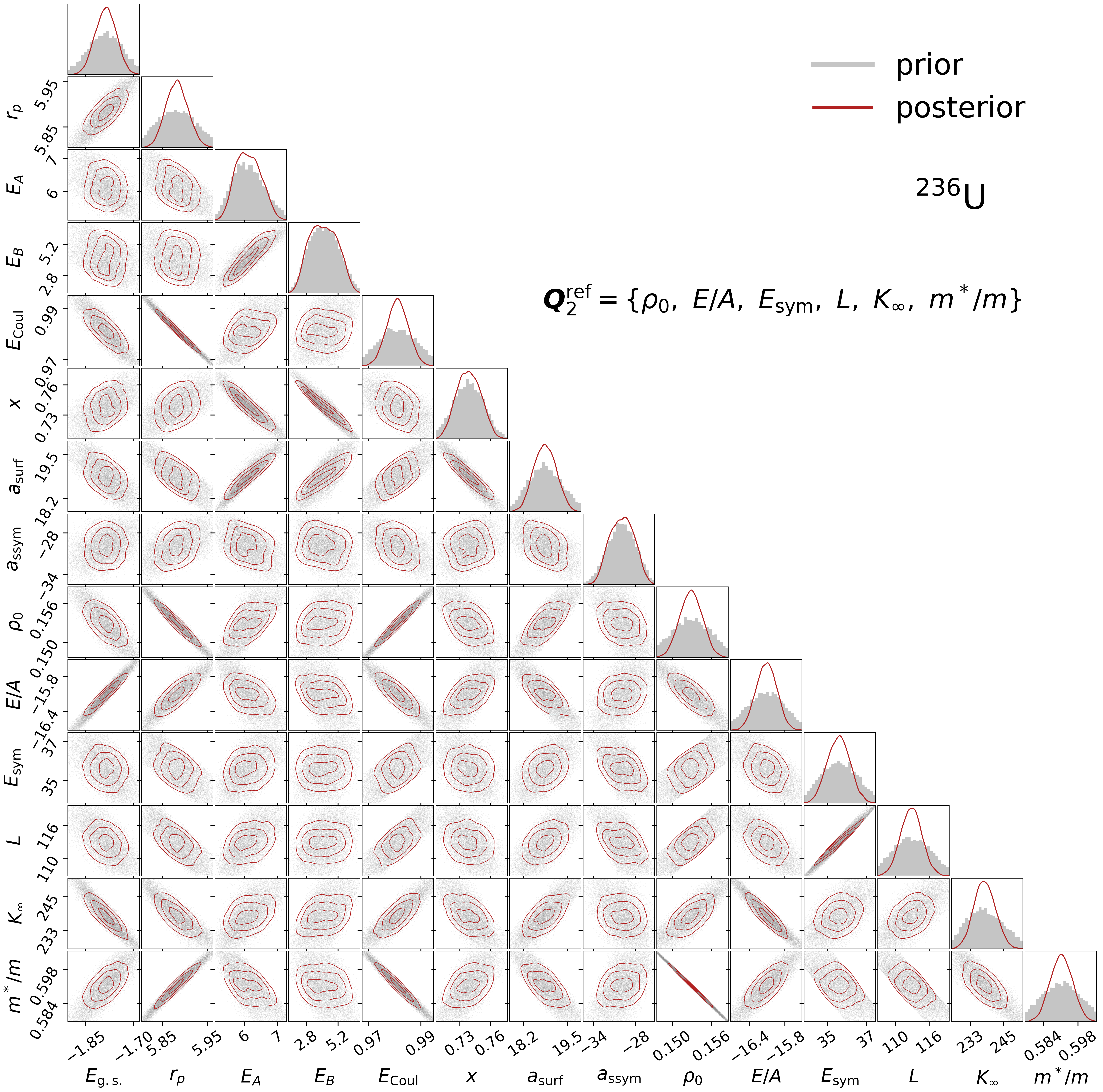}
    \caption{Same as Fig.~\ref{fig-2}, but for the posterior distribution constrained by $\vec{Q}^{\rm ref}_2$.}
    \label{fig-4}
\end{figure*}

\begin{figure}[!htbp]
    \centering 
    \includegraphics[width=.95\columnwidth]{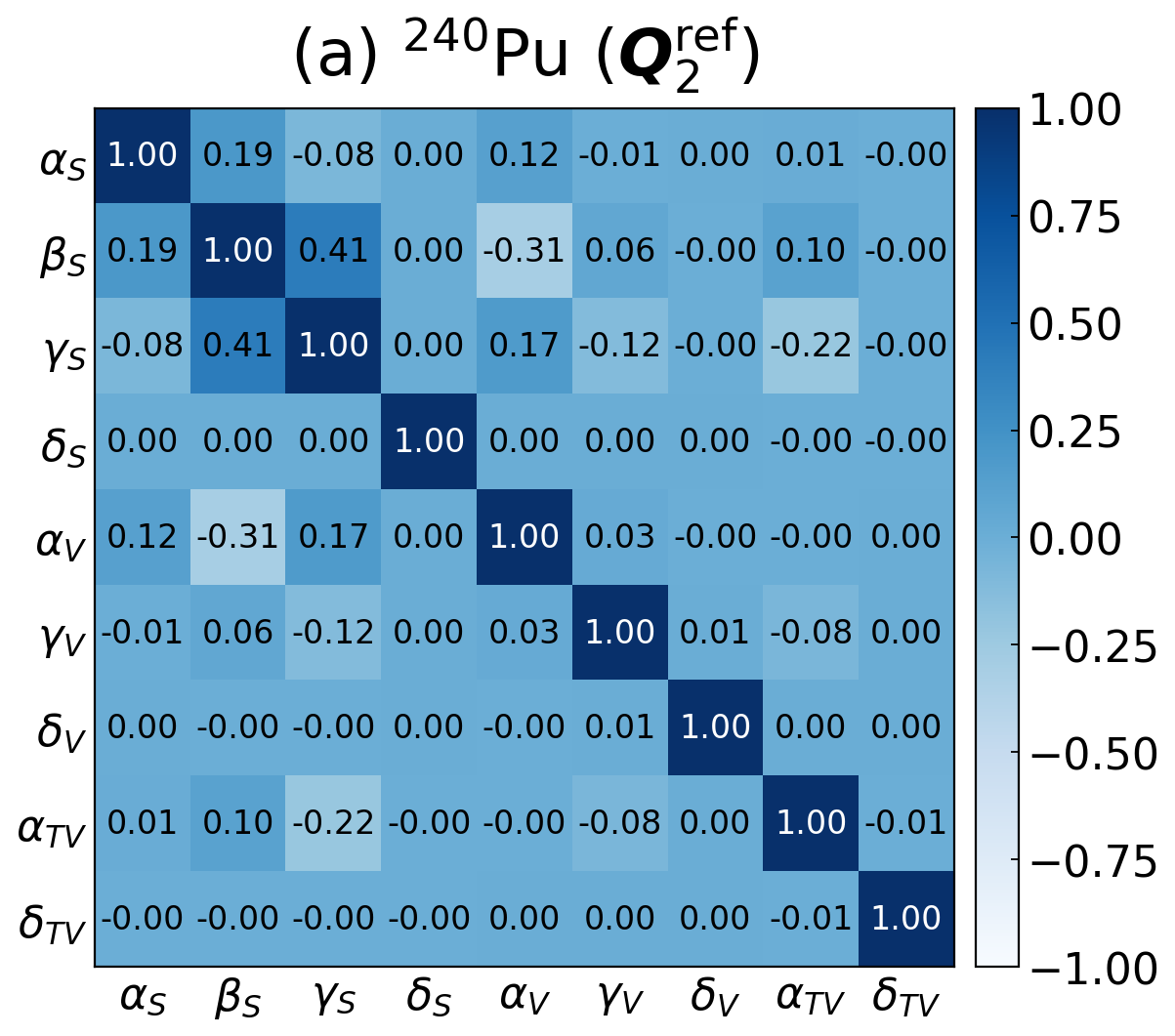}

    \vspace*{1cm}
    
    \includegraphics[width=.95\columnwidth]{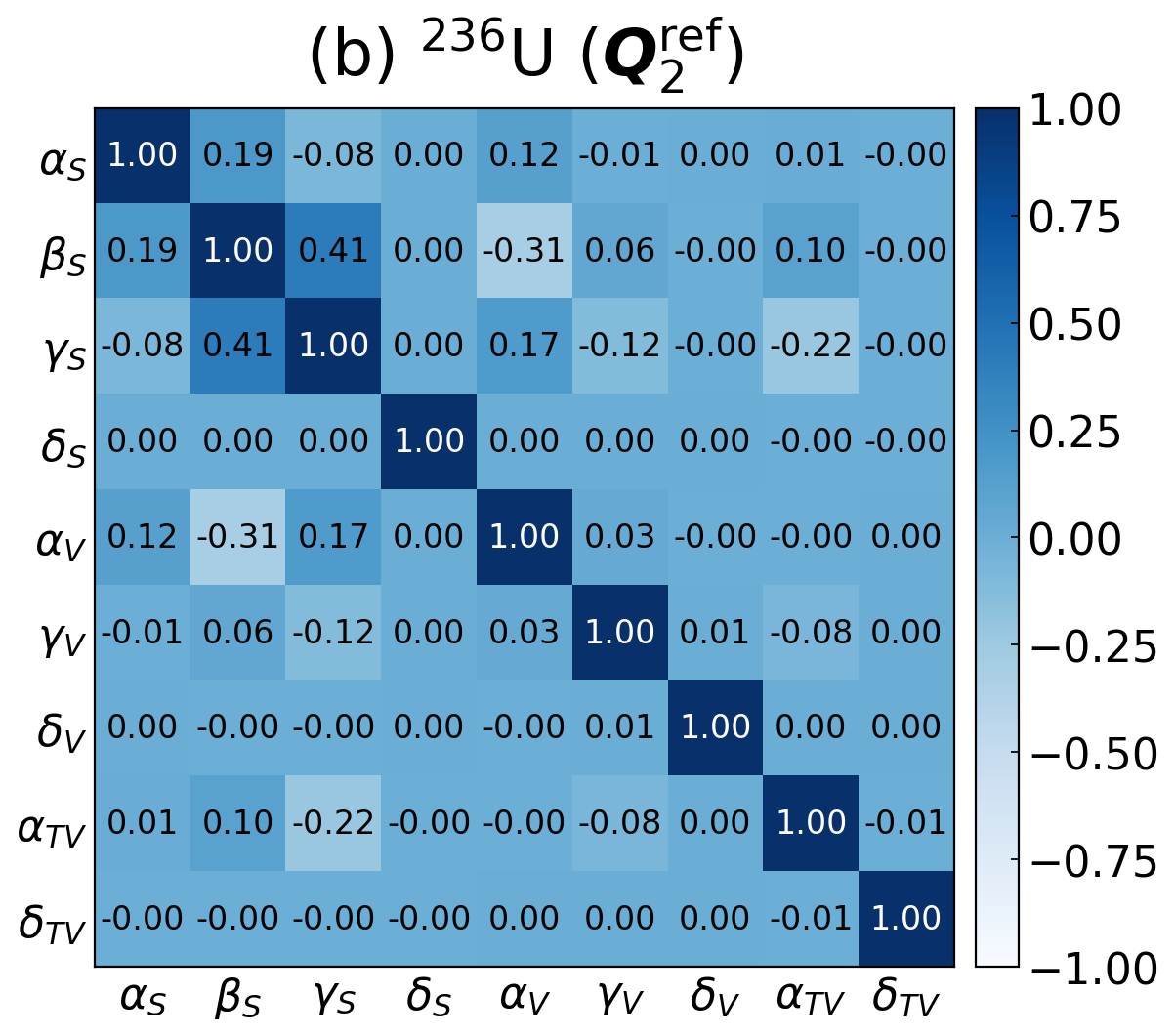}
    \caption{Same as Fig.~\ref{fig:para_corre}, but for the posterior distribution constrained by $\vec{Q}^{\rm ref}_2$.}
    \label{fig:para_corre2}
\end{figure}

\begin{figure}[!htbp]
    \centering
    \includegraphics[width=1.02\columnwidth]{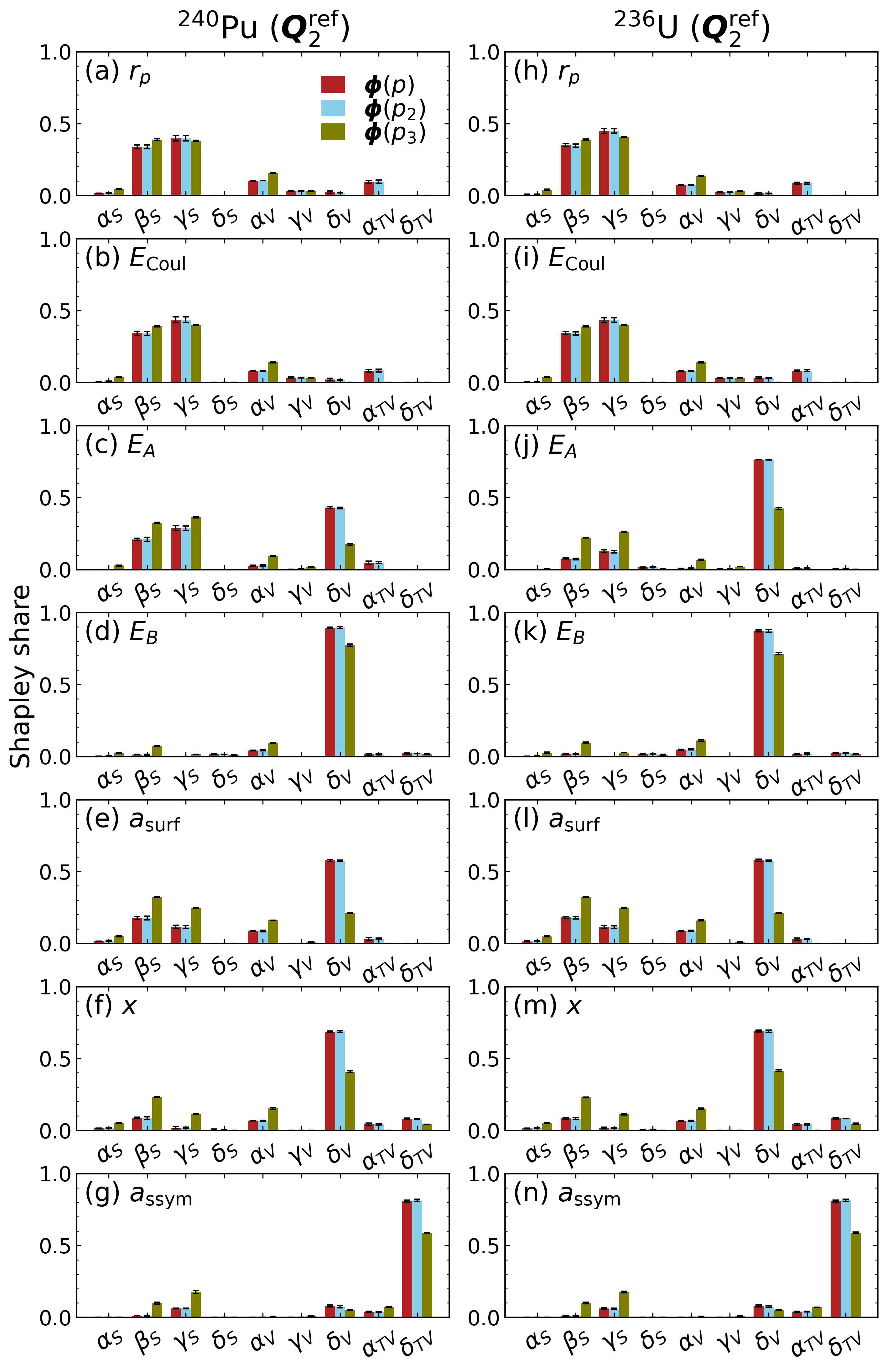}
    \caption{Same as Fig.~4 in the main text, but for the posterior distribution constrained by $\vec{Q}^{\rm ref}_2$.}
    \label{fig:nm_sens1}
\end{figure}

\begin{figure}[!htbp]
    \centering
    \includegraphics[width=1.02\columnwidth]{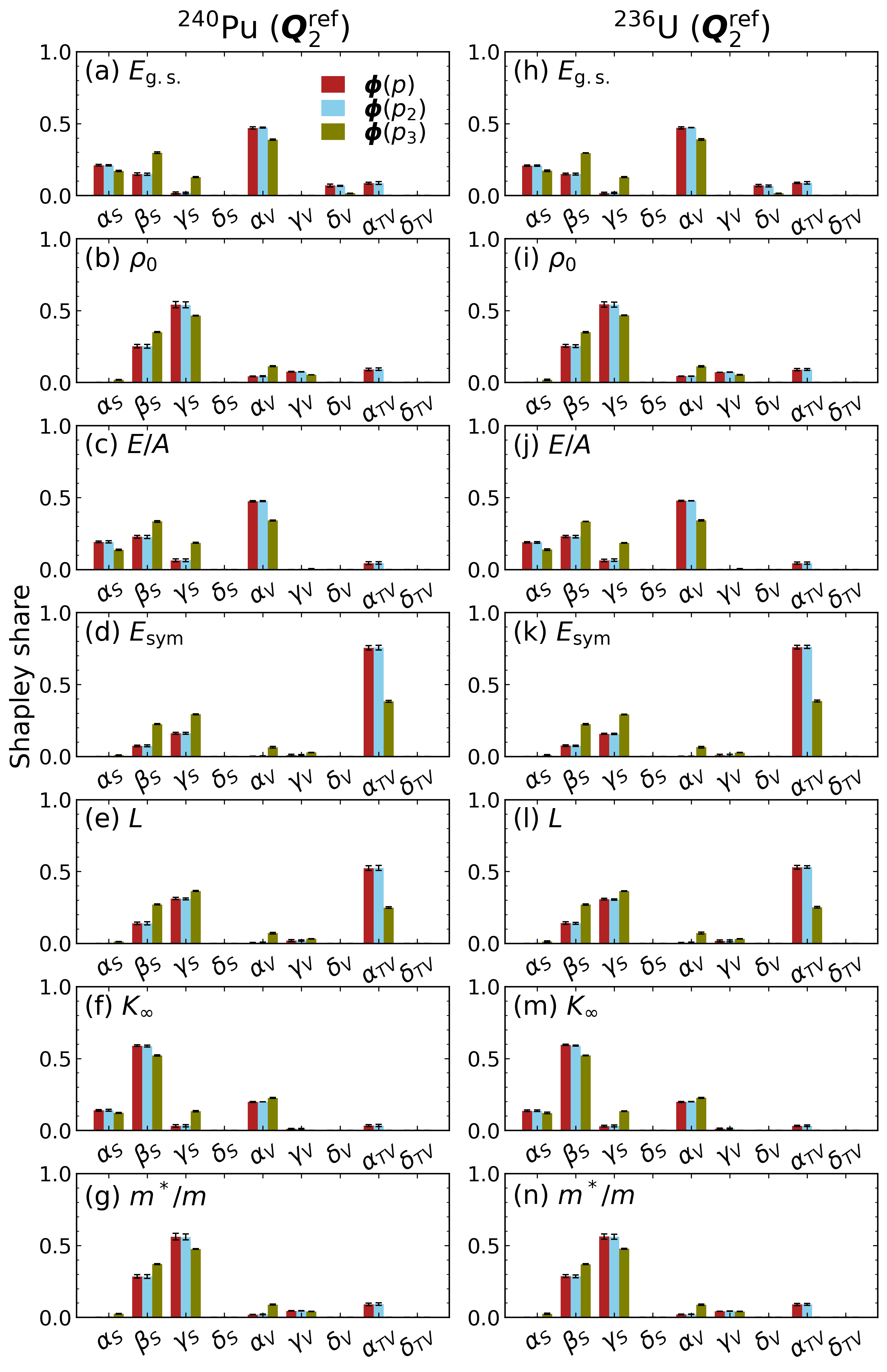}
    \caption{Same as Fig.~\ref{fig:gs_sens}, but for the posterior distribution constrained by $\vec{Q}^{\rm ref}_2$.}
    \label{fig:nm_sens2}
\end{figure}

